\documentclass[journal=nalefd,manuscript=letter,layout=twocolumn]{achemso}
\pdfoutput=1
\usepackage{amsmath,amssymb,amsfonts}
\usepackage[osf]{newpxtext}
\usepackage{newpxmath}
\usepackage[utf8]{inputenx}
\usepackage{graphicx}
\usepackage[dvipsnames]{xcolor}
\usepackage[pdftex]{hyperref}

\DeclareMathAlphabet{\mathcal}{OMS}{cmsy}{m}{n}

\hypersetup{pdfauthor={Thomas Auzelle}, 
    bookmarksnumbered=true, 
    pdftitle={Enhanced radiative efficiency in GaN NWs grown on sputtered TiNx: effects of surface electric fields}, 
    colorlinks, 
    citecolor=blue, 
    linkcolor=blue, 
    urlcolor=blue}

\usepackage{xspace}
\usepackage{soul}
\usepackage{booktabs} 
\usepackage{lineno} 

\newcommand{\degC}{\,$^\circ$C}

\let\oldmaketitle\maketitle
\let\maketitle\relax
\title{Enhanced radiative efficiency in GaN nanowires grown on sputtered TiN$_{\boldsymbol{x}}$: effects of surface electric fields}

\author{T.~Auzelle} 
\affiliation{Paul-Drude-Institut für Festkörperelektronik, Leibniz-Institut im Forschungsverbund Berlin e.\,V., Hausvogteiplatz 5--7, 10117 Berlin, Germany}
\email{auzelle@pdi-berlin.de}
\author{M.~Azadmand}
\affiliation{Paul-Drude-Institut für Festkörperelektronik, Leibniz-Institut im Forschungsverbund Berlin e.\,V., Hausvogteiplatz 5--7, 10117 Berlin, Germany}
\alsoaffiliation{On leave from: L-NESS and Dipartimento di Scienza dei Materiali, Università di Milano-Bicocca, Via R. Cozzi 55, 20125 Milano, Italy}
\author{T.~Flissikowski}
\affiliation{Paul-Drude-Institut für Festkörperelektronik, Leibniz-Institut im Forschungsverbund Berlin e.\,V., Hausvogteiplatz 5--7, 10117 Berlin, Germany}
\author{M.~Ramsteiner}
\affiliation{Paul-Drude-Institut für Festkörperelektronik, Leibniz-Institut im Forschungsverbund Berlin e.\,V., Hausvogteiplatz 5--7, 10117 Berlin, Germany}
\author{K.~Morgenroth}
\affiliation{Paul-Drude-Institut für Festkörperelektronik, Leibniz-Institut im Forschungsverbund Berlin e.\,V., Hausvogteiplatz 5--7, 10117 Berlin, Germany}
\author{C.~Stemmler}
\affiliation{Paul-Drude-Institut für Festkörperelektronik, Leibniz-Institut im Forschungsverbund Berlin e.\,V., Hausvogteiplatz 5--7, 10117 Berlin, Germany}
\author{S.~Fernández-Garrido}
\affiliation{Paul-Drude-Institut für Festkörperelektronik, Leibniz-Institut im Forschungsverbund Berlin e.\,V., Hausvogteiplatz 5--7, 10117 Berlin, Germany}
\alsoaffiliation{Present address: Grupo de Electrónica y Semiconductores, Dpto.\ Física Aplicada, Universidad Autónoma de Madrid, C/ Francisco Tomás y Valiente 7, 28049 Madrid, Spain}
\author{S.~Sanguinetti}
\affiliation{L-NESS and Dipartimento di Scienza dei Materiali, Università di Milano-Bicocca, Via R. Cozzi 55, 20125 Milano, Italy}
\author{H.~T.~Grahn}
\affiliation{Paul-Drude-Institut für Festkörperelektronik, Leibniz-Institut im Forschungsverbund Berlin e.\,V., Hausvogteiplatz 5--7, 10117 Berlin, Germany}
\author{L.~Geelhaar}
\affiliation{Paul-Drude-Institut für Festkörperelektronik, Leibniz-Institut im Forschungsverbund Berlin e.\,V., Hausvogteiplatz 5--7, 10117 Berlin, Germany}
\author{O.~Brandt}
\affiliation{Paul-Drude-Institut für Festkörperelektronik, Leibniz-Institut im Forschungsverbund Berlin e.\,V., Hausvogteiplatz 5--7, 10117 Berlin, Germany}

\begin{document}
\twocolumn[
\begin{@twocolumnfalse}
    \oldmaketitle
\begin{abstract}
	GaN nanowires grown by molecular beam epitaxy generally suffer from dominant nonradiative recombination, which is believed to originate from point defects. To suppress the formation of these defects, we explore the synthesis of GaN nanowires at temperatures up to 915\degC{} enabled by the use of thermally stable TiN$_x$/Al$_2$O$_3$ substrates. These samples exhibit indeed bound exciton decay times approaching those measured for state-of-the-art bulk GaN. However, the decay time is not correlated with the growth temperature, but rather with the nanowire diameter. The inverse dependence of the decay time on diameter suggests that the nonradiative process in GaN nanowires is not controlled by the defect density, but by the field ionization of excitons in the radial electric field caused by surface band bending. We propose a unified mechanism accounting for nonradiative recombination in GaN nanowires of arbitrary diameter.
\end{abstract}
\end{@twocolumnfalse}
]

The recombination dynamics in GaN nanowires (NWs) synthesized by molecular beam epitaxy (MBE) is governed by nonradiative processes even at cryogenic temperatures. In fact, the typical decay times of the donor-bound exciton line in low-temperature photoluminescence (PL) spectra of GaN NWs with diameters between $30$ and $200$\,nm range from $0.1$ to $0.25$\,ns,\cite{Calleja2000,Corfdir2009,Hauswald2014a,Wolz2015} i.\,e., they are significantly shorter than the radiative lifetime of the bound exciton state in bulk GaN of at least $1$\,ns.\cite{Monemar2010} Obvious candidates for nonradiative recombination are the NW surface \cite{gorgis2012} and dislocations generated by NW coalescence.\cite{Consonni2009,Grossklaus2013,Fan2014,Kaganer2016b}
However, \citet{Hauswald2014a} showed that the decay time does not exhibit any obvious trend with either the surface-to-volume ratio or the coalescence degree and thus suggested that the nonradiative channel originates from point defects. This view was supported by subsequent work of \citet{Zettler2016}, who observed an \emph{increase} of the decay time from $0.14$ to $0.36$\,ns upon \emph{thinning} NWs by partial thermal decomposition of a NW ensemble. This striking result was tentatively attributed to a reduction of the density of native point defects by the high-temperature ($920$\degC) annealing of the thinned NWs obtained by sublimation.

The optimum growth temperature for minimizing the point defect density in solids is a trade-off between the limited kinetics at low temperatures and the increasing entropy at high temperatures and is expected to be close to half the melting point of the material ($\approx\!\!1100$\degC{} for GaN).\cite{Burton1951,Ishizaka1994}  Accordingly, \citet{Zettler2015b} proposed to push the MBE growth regime for GaN NWs to far higher growth temperatures ($\gg\!800$\degC{}). However, the authors only reached a maximum temperature of $875$\degC{}, at which Ga-induced melt-back etching of the Si($111$) substrates was observed to set in, resulting in a substantial incorporation of Si into the GaN NWs. For GaN NW growth at even higher temperatures, a thermally and chemically more robust substrate is required. \citet{Wolz2015} demonstrated the use of $\updelta$-TiN($111$) films obtained by surface nitridation of a thick Ti layer deposited onto $\upalpha$-Al$_2$O$_3$($0001$), but this layer was found to react with both the impinging Ga flux and the substrate, with the latter reaction inducing the incorporation of O into the GaN NWs at high growth temperatures.\cite{Calabrese2019} Other groups have used Ti films on different substrates or thick Ti foils, but NWs always formed on a superficial TiN film obtained by nitridation prior to growth,\cite{Sarwar2015,Zhao2016,Calabrese2016,Ramesh2019} leaving the substrate susceptible to reactions with Ga. Note that this interest in TiN as a substrate has various reasons: besides its refractory properties, it is metallic in nature, allowing the synthesis of hybrid metal/semiconductor structures that are attractive for several application areas.\cite{Choi2011,Naik2011,Li2014,May2016,Lu2017}

In this Letter, we set out to synthesize GaN NWs at unprecedentedly high temperatures and to thus find out whether exciton lifetimes approaching those achieved in bulk-like GaN produced by high-temperature deposition techniques can be reached. For this purpose, direct sputtering of TiN$_x$ on Al$_2$O$_3$ is chosen as a new approach for preparing the GaN NW substrate. In contrast to Ti, TiN is stable against Al$_2$O$_3$ up to temperatures far exceeding those usable for the synthesis of GaN in MBE. The TiN$_x$/Al$_2$O$_3$ substrates allow us to fabricate GaN NW ensembles in a temperature range of $855 - 915$\degC{}. The bound exciton lifetimes of these ensembles are longer than any values reported previously for NWs with diameters between 30 and 200~nm, but do not exhibit the expected trend to get longer with increasing growth temperature. Instead, we observe a correlation with the mean diameter of the ensemble, pointing toward exciton dissociation in the radial electric fields in the NWs as the actual mechanism controlling the effective nonradiative lifetime. Indeed, this mechanism is shown to provide a unified frame to quantitatively understand the exciton lifetimes in GaN NWs of arbitrary diameter.    

\begin{table*}
\caption{Parameters for the synthesis of TiN$_x$ films and GaN NWs. A ramp between two values is indicated by $\nearrow$ and $\searrow$. \label{tab:growth}}
\begin{tabular}{lcccc}
    \toprule
    Samples                 						& \textcolor{blue}{A}   &\textcolor{Green}{B}       &\textcolor{orange}{C}        	&\textcolor{red}{D} \\ 
    \midrule              
\multicolumn{5}{l}{\textbf{TiN$_x$ films}}\\
    Ar flux (sccm)              				& $13.5$  				& $13.5$   					& $13.5$  						&  $12\nearrow12.8$ \\
    N$_2$ flux (sccm)          				& $1.5$  					& $1.5$    					& $1.5$  							&  $3\searrow2.2$   \\
    Bias (V)               							    & $100$  					& $100$ 						& $100$ 	 						&  $0$            	\\
\midrule
\multicolumn{5}{l}{\textbf{GaN NWs}}\\	
	Substrate temperature ($^\circ{}$C) 					& $855$ 					& $855\nearrow900$ 			& $855\nearrow915$				& $855$    			\\
	Ga/N flux ratio            						& $0.4$	 				& $0.4$						& $0.4\nearrow1.3$  			& $0.4$      			\\    
	Growth time (min)          						& $90$ 					& $120$						& $220$     						& $120$                \\
	\bottomrule 
 \end{tabular}
\end{table*}

Analogously to GaN NW growth on Si($111$) substrates,\cite{Garrido2015} the nucleation of GaN NWs on TiN$_x$ is preceded by an incubation time during which no stable GaN nuclei form on the surface. On stoichiometric TiN, this incubation time exceeds $1$\,h already at a moderate substrate temperature of $800$\degC{} and a Ga flux as large as $1.2\times10^{15}$\,cm$^{-2}$\,s$^{-1}$ (III/V ratio of $1$). At higher temperatures, no GaN NWs form for any practical time span. The incubation time is found to drastically decrease on TiN$_x$ films with a certain N deficiency. Specifically on TiN$_{0.88}$, the incubation time is shorter than $5$\,min at a substrate temperature of $855$\degC{} and with a Ga flux of $0.5\times10^{15}$\,cm$^{-2}$\,s$^{-1}$, which we adopt as the conditions for NW nucleation for the present samples A--D. The substrate temperature is increased  during the NW elongation stage \cite{Zettler2015a} for samples B and C. The enhanced thermal decomposition of GaN at $915$\degC{} is partially compensated by a larger Ga flux for sample C. All growth parameters for samples A--D are summarized in Table\,\ref{tab:growth}.

\begin{figure*}[t]
\includegraphics[width = \linewidth]{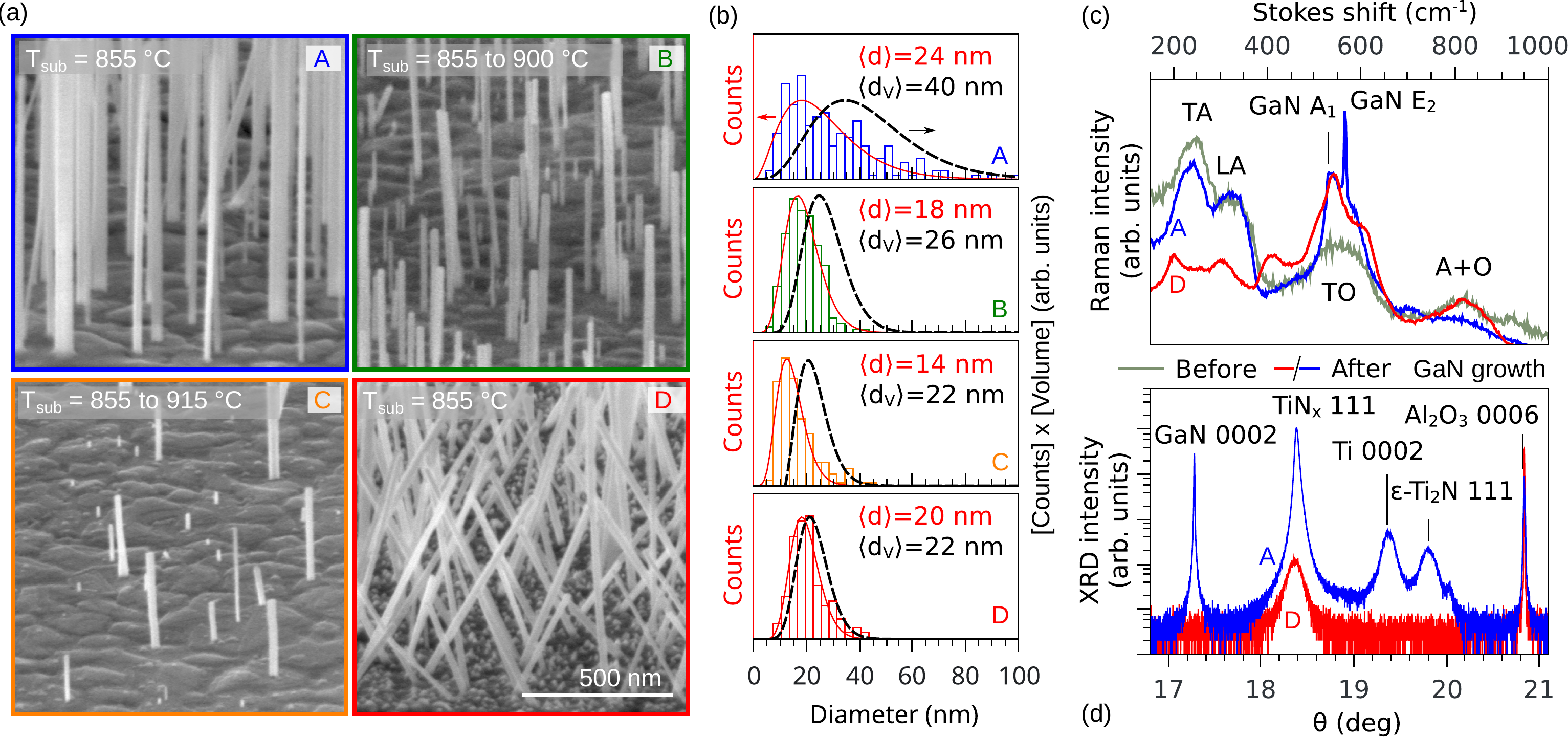}%
\caption{(a) Bird's eye view secondary electron micrographs of the GaN NW ensembles grown on sputtered TiN$_x$. The scale bar for sample D applies to all micrographs. (b) Diameter distributions of the NW ensembles extracted from bird's eye view secondary electron micrographs. The solid lines are fits of the histograms with a shifted $\Gamma$ distribution. The dashed lines show the volume-weighted distributions. The mean diameters of these distributions are indicated in the figure. (c) Raman spectra and (d) symmetric $\upomega/2\uptheta$  scans of samples A (blue line) and D (red line). A representative Raman spectrum of an as-sputtered TiN$_x$ film is also shown as a gray line in (c). The features observed are due to the transverse acoustic (TA), longitudinal acoustic (LA) and transverse optical (TO) modes as well as second-order scattering processes including acoustic and optical phonons (A+O).\cite{Spengler1978}
\label{fig:NWs}}
\end{figure*}

Figure~\ref{fig:NWs}(a) shows bird's eye view secondary electron micrographs of the NW ensembles A to D. The increase in temperature from sample A to B to C results in shorter NWs and a reduction of the NW density from $2.5 \times 10^9$ to less than $1.0 \times 10^{9}$\,cm$^{-2}$. The NWs are vertical due to a strict epitaxial orientation relationship with the monocrystalline TiN$_x$($111$) film: the width of the out-of-plane and in-plane orientation distributions as measured by x-ray diffraction (XRD) $\upomega$ and $\upphi$ scans recorded for sample A (not shown here) are below $0.7^\circ$. The nearly random orientation of NWs observed for sample D is again caused by an epitaxial alignment, but in this case the TiN film is polycrystalline due to the absence of a substrate bias during sputtering.

Figure \ref{fig:NWs}(b) depicts the distributions of diameter $d$ for samples A--D. The histograms are fit by a shifted $\Gamma$ distribution yielding the mean diameter $\langle d\rangle$. Also shown is the diameter distribution weighted by the NW volume, which is the relevant quantity for experimental techniques returning a signal proportional to the amount of probed material such as PL spectroscopy. Evidently, higher growth temperatures do not only decrease the NW density (see above), but also result in smaller mean diameters and narrower diameter distributions as a result of a reduced NW coalescence.\cite{Brandt2014,vanTreeck2018} Ensemble D is an exception in that it features a much narrower diameter distribution compared to sample A nominally grown under the same conditions. This difference stems from the random orientation of the NWs in sample D, which effectively inhibits their coalescence by bundling.\cite{Kaganer2016a} 
Ensemble D is thus essentially coalescence-free despite a relatively high NW density ($5\times 10^9$\,cm$^{-2}$).  

Figures~\ref{fig:NWs}(c) and \ref{fig:NWs}(d) show Raman spectra and radial XRD scans, respectively, of samples A and D together with a representative Raman spectrum of an as-sputtered TiN$_x$ film. For sample A, modes from the TiN$_x$ film\cite{Spengler1978} and from the GaN NWs (A$_1$ and E$_2$\cite{Davydov1997}) are visible. The transverse acoustic (TA) mode of the TiN$_x$ film shifts from $248$ before to $242$\,cm$^{-1}$ after the NW growth, evidencing a reduction of the concentration of N vacancies due to the prolonged annealing during growth. The same spectral position, corresponding to $x=0.93$,\cite{Spengler1978} is observed for samples B and C after growth. In the XRD profile of sample A (as well as for B and C), we detect a strong and sharp TiN $111$ reflection as for stoichiometric TiN,\cite{Azadmand2020} but also two weak additional reflections. The angular position, intensity, and width of these reflections indicate that our TiN$_x$ films contain nanoscopic inclusions of Ti and $\upvarepsilon$-Ti$_2$N with a total volume on the order of 1\%.\cite{Pecz1995} 

Sample D is different in all these regards. First, the TA mode shifts from $248$ to $200$\,cm$^{-1}$ after growth, the value for stoichiometric TiN.\cite{Spengler1978} Clearly, the polycrystalline nature of this TiN film expedites the annihilation of N vacancies. Second, the TiN 111 reflection for sample D is weak due to the polycrystalline nature of this film, and the NWs are barely detected because of their large orientational spread. 

\begin{figure*}
\includegraphics[width = \linewidth]{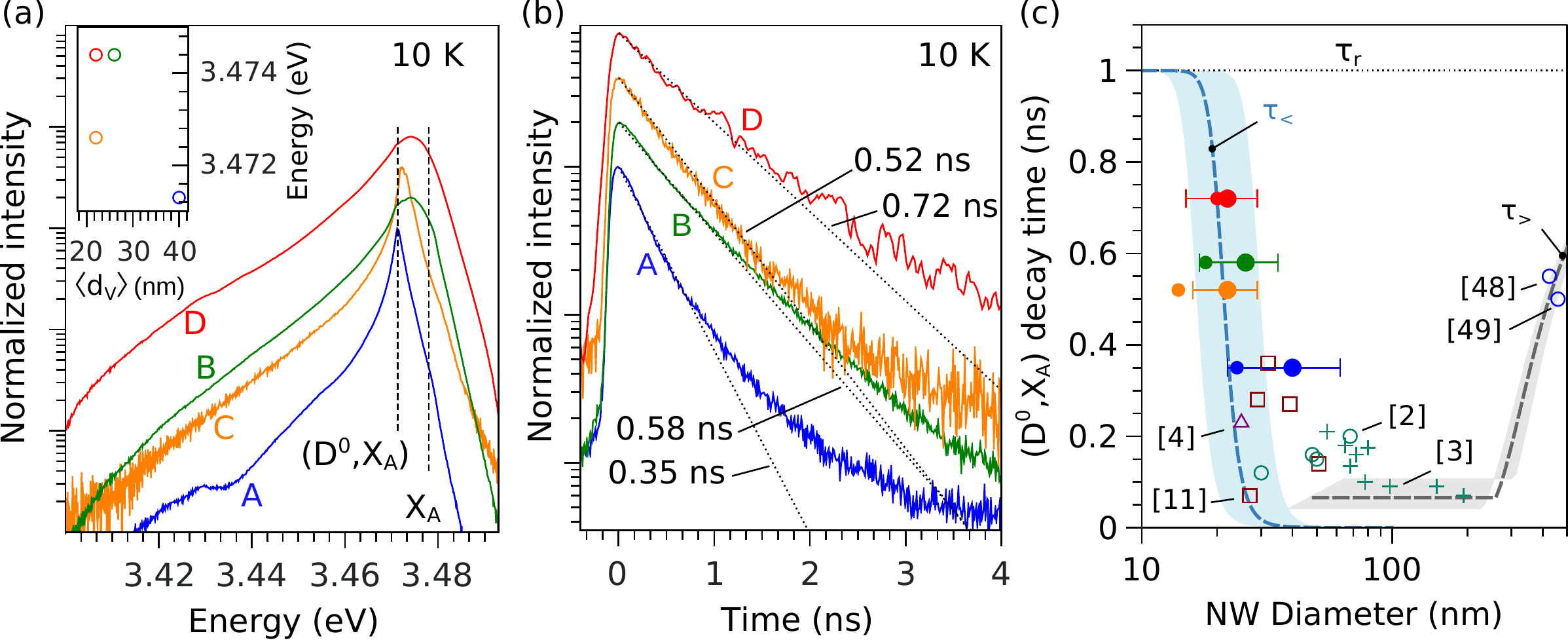}%
\caption{(a) cw-PL spectra and (b) band-edge PL transients integrated ($\pm 15$~meV) over the $(D^0,X_\text{A})$ line acquired at 10\,K from samples A--D. The inset in (a) shows the peak energy vs.\ the volume-weighted mean diameter of the ensemble.  The dotted lines in (b) are monoexponential fits of the initial decay times, which are also given in the figure. (c) $(D^0,X_\text{A})$ decay time versus the average NW diameter for samples A--D and samples from other experimental studies. Large filled circles with horizontal error bars correspond to $\langle d_v\rangle$, whereas small filled circles refer to $\langle d\rangle$. Diameters from other publications are not weighted and are overestimated in the particular case of Ref.\citenum{Zettler2016}. $\tau_\text{r}$ is the radiative lifetime of the $(D^0,X_\text{A})$. $\tau_<$ and $\tau_>$ are analytical expressions of the $(D^0,X_\text{A})$ decay time in the occurrence of field ionization as the dominant nonradiative recombination channel. They are calculated in the limit of \emph{thin} and \emph{thick} NWs, respectively. The exact range of validity for these expressions is calculated in section III of the Supporting Information. A good agreement with experimental values is obtained for a donor concentration of $7 \pm 2 \times 10^{16}$\,cm$^{-3}$ and $L_\text{D} = 32$\,nm. The shaded areas reflect the uncertainty in the donor concentration.
\label{fig:PL}}
\end{figure*}

Figure~\ref{fig:PL}(a) displays the PL spectra of our samples recorded under continuous-wave (cw) excitation at 10\,K. The energy of the emission band identifies the transition as being due to the radiative decay of the donor-bound exciton $(D^0,X_\text{A})$. However, the emission energy is diameter dependent and not identical to that of bulk GaN as usually observed for NWs with larger diameters, but exhibits on average a blueshift with decreasing diameter (see the inset). Several mechanisms have been identified that can shift the transition energy for thin NWs, namely, surface donor-bound excitons,\cite{Corfdir2009,Brandt2010,Corfdir2014a} surface stress,\cite{Calabrese2020} and dielectric confinement.\cite{Zettler2016} Since the shift depends on diameter, all of these mechanisms result in a broadening of the ensemble spectra as primarily observed here for samples B and D. In addition, NWs with a very small volume may not contain even a single donor \cite{Pfuller2010a} so that the free-exciton transition may appear to be strongly enhanced in low-temperature ensemble spectra as seen here particularly for sample B.

PL intensity transients of the $(D^0,X_\text{A})$ line obtained upon pulsed excitation at 10\,K are shown in Figure~\ref{fig:PL}(b). The biexponential decay was shown to arise from a coupling of the bound exciton states \cite{Hauswald2013} with the initial decay time reflecting the actual lifetime of the $(D^0,X_\text{A})$ state \cite{Hauswald2013} that so far remained mostly in the range from $0.1$ to $0.25$\,ns.\cite{Hauswald2014a} For the present samples, exceptionally long initial decay times between $0.35$ and $0.72$\,ns are observed, but not the expected correlation with the growth temperature. In particular, sample D features a decay time twice longer than the one of sample A grown at the exact same temperature. A comparatively clear trend is evident instead when correlating the decay time with the volume-weighted mean diameter $\langle d_v\rangle$, as shown in Figure~\ref{fig:PL}(c), suggesting that the decay time \emph{increases} with \emph{decreasing} diameter. As a matter of fact, such a behavior was observed previously by \citet{Zettler2016} and interpreted as an effect of NW annealing on the point defect density. Our results suggest that the key parameter is the NW diameter itself, calling for a reevaluation of the mechanism responsible for the nonradiative exciton decay in GaN NWs. 

\begin{figure*}
\includegraphics[width = \linewidth]{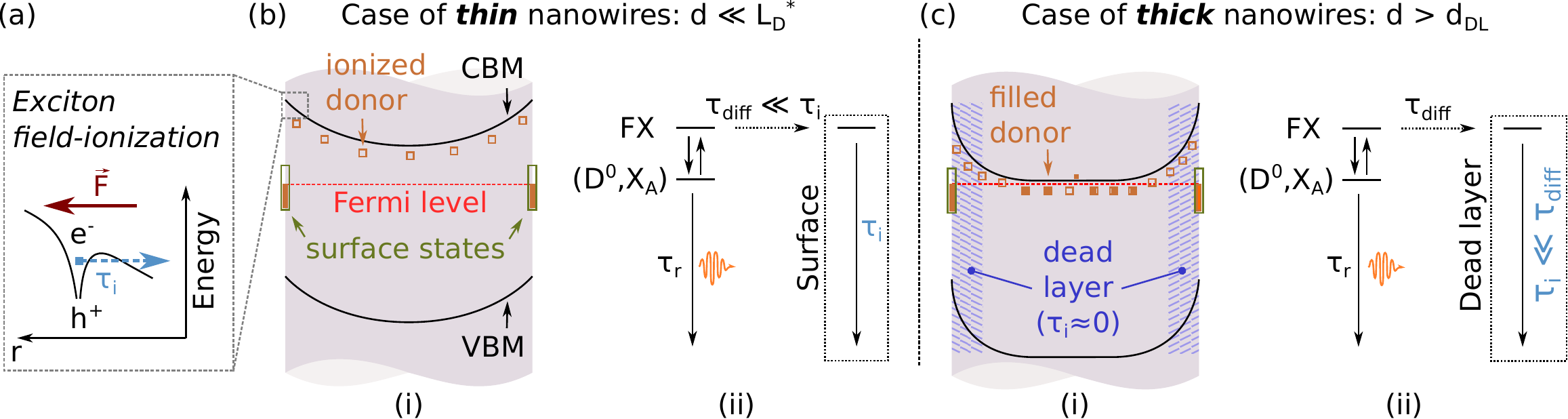}%
\caption{(a) Illustration of exciton dissociation in an electric field: the electron tunnels out of the tilted Coulomb potential created by the hole in a characteristic time $\tau_\text{i}$ (blue dotted arrow). The panels (i) in (b) and (c) show the conduction (CBM) and valence band edges (VBM) across the NW in the two limiting cases of \emph{thin} and \emph{thick} NWs, respectively, corresponding to $d \ll L_\text{D}^*$ and $d > d_\text{DL}$. $d$ is the NW diameter, $L_\text{D}^* = (D \tau_\text{i}^\text{surf})^{1/2}$  the exciton diffusion length in the characteristic time $\tau_\text{i}^\text{surf}$ for field ionization of the exciton at the NW surface, and $d_\text{DL}$ the diameter above which an exciton dead layer develops below the surface. The panels (ii) in (b) and (c) depict the competing decay channels for the exciton in \emph{thin} and \emph{thick} NWs, respectively.
\label{fig:Mechanism}}
\end{figure*}

At this point, it is important to recall that surface band bending occurs in GaN NWs due to the presence of surface states at their $\{1\bar{1}00\}$ sidewall facets.\cite{Calarco2005,Portz2018} A reduction of the surface band bending was observed to enhance the intensity of the $(D^0,X_\text{A})$ line,\cite{Pfuller2010b} suggesting that the electric field associated with the band bending triggers the nonradiative decay of the $(D^0,X_\text{A})$ state, while the more strongly bound excitons remain unaffected.\cite{Corfdir2014b} These findings can be understood when taking into account that the nonradiative decay occurs via the free exciton.\cite{Hauswald2014a} The comparatively low binding energy of the $(D^0,X_\text{A})$ state ($\approx 7$\,meV) results in a strong coupling with the free exciton already at 10\,K as revealed by the fact that both share a common lifetime.\cite{Hauswald2014a,Corfdir2014b} An effective nonradiative decay process can thus be initiated in the vicinity of the surface by the field ionization of the free exciton, i.\,e., the tunneling of an electron out of the Coulomb potential created by the hole,\cite{Blossey1970} as illustrated in Figure~\ref{fig:Mechanism}(a). The free electron and holes are driven apart in the field, with the holes drifting to the surface where they rapidly recombine with filled surface states.\cite{Winnerl2015} Subsequently, the excess electrons are captured by the empty states once more. 

For a quantitative comparison with experiments, we calculate the decay times of the $(D^0,X_\text{A})$ state in the presence of surface electric fields for the two limiting cases of \emph{thin} and \emph{thick} NWs. 
The strength of the surface electric fields is determined in the approximation of a continuous and homogeneous charge distribution with a background $n$-type doping.\cite{Corfdir2014a} The field ionization time $\tau_\text{i}$ for an exciton in an electric field $F$ is derived from the ionization probability (per unit time) of the hydrogen atom \cite{Yamabe1977,Banavar1979} following Ref.\citenum{Kaganer2018}. The exponential relationship results in a steep dependence of $\tau_\text{i}$ on $F$. Besides field ionization of the free exciton, radiative recombination of the $(D^0,X_\text{A})$ complex with a lifetime $\tau_\text{r} = 1$\,ns is considered. These calculations are discussed in detail in sections I and II of the Supporting Information. The analytical expressions for the effective decay times $\tau_<$ and $\tau_>$ obtained for the limiting cases of \emph{thin} and \emph{thick} NWs, respectively, are given below.

In \emph{thin} NWs, all electrons from shallow donors transfer to surface states, the NW is entirely depleted, and the surface band bending thus extends through the whole NW as depicted in panel (i) of Figure~\ref{fig:Mechanism}(b). The maximum electric field occurs at the NW sidewall facets and increases linearly with NW diameter and doping density. Correspondingly, the field ionization rate of the free exciton is maximal at the surface and increases for larger diameter and doping density. Panel (ii) in Figure~\ref{fig:Mechanism}(b) shows the two competing decay channels determining $\tau_<$, the decay time of the $(D^0,X_\text{A})$ state in \emph{thin} NWs : (1) the radiative recombination and (2) the nonradiative recombination via diffusion of the free exciton to the NW surface where it is ionized. For sufficiently thin NWs, the diffusion time of the free exciton is negligible compared to the field ionization time, i.\,e.\ field ionization represents the rate-limiting step for nonradiative recombination. This condition is met for $d \ll L_\text{D}^*$, which sets the upper bound of \emph{thin} NWs. Here, $L_\text{D}^* = (D \tau_\text{i}^\text{surf} )^{1/2}$ represents the diffusion length of the free exciton in the characteristic time $\tau_\text{i}^\text{surf}$ for field ionization of the exciton at the NW surface and with the free exciton diffusivity $D$. Then, in \emph{thin} NWs $\tau_<$ is given by:
\begin{equation}
 \frac{1}{\tau_<} = \frac{1}{\tau_\text{r}} + \frac{1}{\tau_\text{i}^\text{surf}}\,.
\end{equation}

Figure~\ref{fig:PL}(c) compares the measured $(D^0,X_\text{A})$ decay times for samples investigated in the present work (solid circles) with previous studies \cite{Schlager2008,Corfdir2009,Schlager2011,Hauswald2014a,Wolz2015,Zettler2016} (open symbols). The experimental data exhibit a $\cup$-shaped dependence on $d$: the measured decay times first steeply decrease with an increase of the diameter from 20 to 30\,nm, enter a plateau for diameters between $30$ and $200$\,nm, and then increase for even larger diameters. Certainly the most striking finding is the systematically \emph{increasing} decay time with \emph{decreasing} diameter for very thin NWs, which is exactly the opposite behavior expected for a recombination process involving the surface. Considering the two-step process discussed above, which requires excitons to dissociate prior to surface recombination, this behavior is actually the one we should expect. Figure~\ref{fig:PL}(c) shows the calculated decay time  $\tau_<$ assuming a residual doping level of $N_\text{D} = (7 \pm 2) \times 10^{16}$\,cm$^{-3}$, derived in the limit of thin NWs, which is seen to quantitatively reproduce the trend observed for the decay times of NW ensembles with an average diameter below $ \approx 30$\,nm.  
In ultrathin NWs ($d \ll 20$\,nm), the decay time is predicted to be no longer affected by the electric field, but to represent the actual lifetime of the $(D^0,X_\text{A})$ state. The $0.72$\,ns decay time for sample D approaches the longest values measured for state-of-the-art GaN layers ($\approx 1$\,ns), and even thinner NWs may thus help to settle the question whether this value represents the true radiative lifetime.

Panel (i) in Figure~\ref{fig:Mechanism}(c) depicts the band bending in a very \emph{thick} NW with a nondepleted core. Flat band conditions are established in the NW core, and electric fields exist only in the depleted area below the surface. Yet, for a doping density of $N_\text{D} = 7 \times 10^{16}$\,cm$^{-3}$, the strength of the surface electric field exceeds $14$\,kV/cm (corresponding to essentially instantaneous ionization; see Supporting Information) up to a distance of about $100$\,nm below the surface. This region corresponds to a dead layer \cite{Schultheis1984,Shokhovets2003} since field ionization of the exciton occurs in less than a ps and radiative recombination can thus only occur in the NW core. For the doping density assumed, this dead layer starts to appear in (still fully depleted) NWs with a diameter above $d_\text{DL} \approx 40$\,nm. Panel (ii) in Figure~\ref{fig:Mechanism}(b) shows that the nonradiative recombination in \emph{thick} NWs is limited by the diffusion of the free exciton from the NW core to the edge of the dead layer where the recombination velocity becomes essentially infinite. The decay time $\tau_>$ is given by:
\begin{equation}
    \tau_> = \tau_\text{r} \frac{I_2  (d_\text{c}/L_\text{D} )}{ I_0 (d_\text{c}/L_\text{D}  )}\,,
\end{equation}
where $I_n$ denotes the modified Bessel function of the first order, $d_\text{c}$ the NW diameter after subtraction of the dead layer, and $L_\text{D}$ the exciton diffusion length.

The calculated decay time $\tau_>$ for the $(D^0,X_\text{A})$ state in thick NWs is shown in Figure~\ref{fig:PL}(c) together with the available experimental data. Assuming, as before, a doping level of $N_\text{D} = (7 \pm 2) \times 10^{16}$\,cm$^{-3}$ and a diffusion length $L_\text{D}$ of 30\,nm, both the plateau between 50 and 200\,nm as well as the linear increase for even larger NW diameters are reproduced. The plateau observed for depleted NWs ($d < 260$\,nm) simply reflects the fact that the volume of the radiative core does not depend on NW diameter in this regime. For even larger diameters and the progressive return to flat-band conditions, the diameter of the radiative core increases linearly with the NW diameter, and the nonradiative decay rate thus scales inversely with the NW diameter, a feature reminiscent of conventional surface recombination.\cite{Schlager2008} 


The two-step process detailed above thus provides a framework for a unified understanding of the exciton decay in GaN NWs of arbitrary diameter, but also accounts for other experimental results. For example, the nonradiative lifetime in GaN NWs was observed to be essentially independent of both temperature (between 10 and 300\,K) and excitation density (over four orders of magnitude),\cite{Hauswald2014a} which cannot be understood within the conventional framework of defect-mediated nonradiative recombination. When considering field ionization of excitons as the mechanism inducing the nonradiative decay of excitons, both findings are easily understood: first, the band bending in the NWs originates from the charge transfer from the bulk into surface states and is thus essentially independent of temperature. Equally important, the exciton binding energy in GaN is sufficiently high for excitons still dominating the spontaneous emission at room temperature. Second, the number of surface states in a NW is very high compared to that of point defects in its bulk, rendering the nonradiative channel hard to saturate. 

Having identified the origin for the highly effective nonradiative channel for GaN NWs, the remedy becomes obvious. Since the nonradiative decay is mediated by the field ionization of excitons, the radial electric fields should be eliminated or at least reduced. This task can be achieved by either reducing unintentional doping or, perhaps more practically, by a suitable surface passivation.\cite{Zhao2015,Varadhan2017,Hofmann2017,Wang2017,Latzel2017,Wong2018,Auzelle2020} On the contrary, the growth of a thin shell with a larger band gap\cite{Hetzl2017} [e.g.\, made of (Al,Ga)N] is generally inadequate to cancel the surface electric fields. We stress that the mechanism described in this work is not restricted to GaN NWs, but applies to nano- and microstructures in general. In fact, the occurrence of electric fields (either due to charge transfer or strain) is hard to avoid in these structures,\cite{lewyanvoon_2011,Geijselaers2018} and materials with lower exciton binding energy such as, for example, GaAs, may be affected by these fields even more severely. Field ionization may thus very well limit the performance of devices such as micro-light-emitting diodes and single-photon emitters regardless of the specific material. Last but not least, the exceptionally long exciton lifetime obtained here for the thinnest NWs documents the high crystalline quality achievable for GaN on sputtered TiN. This highly stable, refractory substrate thus provides a robust platform for the fabrication of hybrid metal/semiconductor structures up to very high temperatures.\cite{Azadmand2020}
 
\section{Methods}
\subsection{TiN$_x$ synthesis}

The $\updelta$-TiN$_x$ films are deposited onto $\upalpha$-Al$_2$O$_3$(0001) wafers by reactive sputtering of a Ti target in a mixed Ar/N$_2$ gas environment. %
The stoichiometry of the films is controlled by tuning the Ar:N$_2$ gas composition,\cite{Logothetidis1996} and the parameters selected are summarized in Table\,\ref{tab:growth}.
During sputtering, the substrate is heated to $300$\degC{} and, unless otherwise mentioned (see Table\,\ref{tab:growth}), biased at $100$\,V relative to the target and chamber walls.\cite{Zgrabik2015} The $700$\,nm-thick sputtered films are subsequently transferred in ultra-high vacuum into the MBE chamber to prevent surface oxidation and contamination.

\subsection{Nanowire growth}

In the MBE chamber, the substrate temperature is measured by an optical pyrometer calibrated with the $7\!\times\!7\!\leftrightarrow\!1\!\times\!1$ reconstruction transition of a clean Si($111$) surface occurring at $860$\degC{}.\cite{Suzuki1993} The emissivity of TiN$_{x\lesssim1}$ is assumed to be equal to that of stoichiometric TiN, namely, $0.21$.\cite{Briggs2017} Prior to GaN growth, the TiN$_x$ film is annealed at $955$\degC{} under a N flux of $1.2\times10^{15}$\,cm$^{-2}$\,s$^{-1}$ for $25$\,min. This annealing step generally results in a bright TiN$_x$ diffraction pattern as observed by reflection high-energy electron diffraction, indicating a well-ordered surface. NW nucleation is subsequently initiated by exposing the TiN$_x$ film to Ga and N fluxes, while ramping the substrate temperature to the desired value. All NW growth runs are performed with a N flux of $1.2\times10^{15}$\,cm$^{-2}$\,s$^{-1}$ using a plasma source. 

\subsection{TiN$_x$ microstructrure}

The phase content of the TiN$_x$ films is analyzed by symmetric triple-axis $\upomega/2\uptheta$ XRD scans acquired with Cu$_{K\alpha_1}$ radiation in a PANanalytical X'Pert Pro MRD\texttrademark\ equipped with a Ge($220$) hybrid monochromator and a Ge($220$) analyzer.
Their stoichiometry is estimated from the first-order TA phonon energy in Raman spectra \cite{Spengler1978} measured in backscattering geometry using a laser wavelength of $473$\,nm. 

\subsection{Nanowire spectroscopy}
The radiative and nonradiative decay of excitons in the GaN NWs is probed by cw-PL and time-resolved PL (TRPL) spectroscopy. In both cases, an excitation wavelength of $325$\,nm is used, with an intensity below $10$\,kW\,cm$^{-2}$ ($10$\,W\,cm$^{-2}$) at $300$\,K ($10$\,K) for the former and $200$\,fs pulses with a fluence below $0.1$\,\textmu J\,cm$^{-2}$ for the latter.

\section{Author information}
Optimization of the TiN$_x$ layers: K.M., C.S, S.F.-G., and M.A.; Nanowire synthesis: M.A., K.M., and T.A.; Spectroscopy: T.F., M.R., M.A., T.A, and O.B., Modeling: T.A. and O.B.; Writting of the manuscript: T.A. and O.B., with contributions of all authors. All authors have given approval to the final version of the manuscript.

\begin{suppinfo}
 Calculation of the electric-field strength in nanowires, calculation of the field ionization rate of excitons, calculation of the width of the exciton dead layer, calculation of the exciton decay rate in \emph{thin} and \emph{thick} nanowires, calculation of the boundary delimiting \emph{thin} and \emph{thick} nanowires.
\end{suppinfo}

\begin{acknowledgement}
The authors thank E. Zallo and P. Corfdir for a critical reading of the manuscript. Funding from the German Bundesministerium für Bildung und Forschung through project No.\ FKZ:13N13662, from Regione Lombardia (Italy) through project COSMITO, and from the Spanish Ram\'on y Cajal program (grant RYC-2016-19509 co-financed by the European Social Fund) are gratefully acknowledged.
\end{acknowledgement}

\bibliography{GaNNWs_Ti-TiN}

\providecommand{\latin}[1]{#1}
\makeatletter
\providecommand{\doi}
  {\begingroup\let\do\@makeother\dospecials
  \catcode`\{=1 \catcode`\}=2 \doi@aux}
\providecommand{\doi@aux}[1]{\endgroup\texttt{#1}}
\makeatother
\providecommand*\mcitethebibliography{\thebibliography}
\csname @ifundefined\endcsname{endmcitethebibliography}
  {\let\endmcitethebibliography\endthebibliography}{}
\begin{mcitethebibliography}{66}
\providecommand*\natexlab[1]{#1}
\providecommand*\mciteSetBstSublistMode[1]{}
\providecommand*\mciteSetBstMaxWidthForm[2]{}
\providecommand*\mciteBstWouldAddEndPuncttrue
  {\def\EndOfBibitem{\unskip.}}
\providecommand*\mciteBstWouldAddEndPunctfalse
  {\let\EndOfBibitem\relax}
\providecommand*\mciteSetBstMidEndSepPunct[3]{}
\providecommand*\mciteSetBstSublistLabelBeginEnd[3]{}
\providecommand*\EndOfBibitem{}
\mciteSetBstSublistMode{f}
\mciteSetBstMaxWidthForm{subitem}{(\alph{mcitesubitemcount})}
\mciteSetBstSublistLabelBeginEnd
  {\mcitemaxwidthsubitemform\space}
  {\relax}
  {\relax}

\bibitem[Calleja \latin{et~al.}(2000)Calleja, S\'anchez-Garc\'{\i}a, S\'anchez,
  Calle, Naranjo, Mu\~noz, Jahn, and Ploog]{Calleja2000}
Calleja,~E.; S\'anchez-Garc\'{\i}a,~M.~A.; S\'anchez,~F.~J.; Calle,~F.;
  Naranjo,~F.~B.; Mu\~noz,~E.; Jahn,~U.; Ploog,~K. Luminescence properties and
  defects in GaN nanocolumns grown by molecular beam epitaxy. \emph{Phys. Rev.
  B} \textbf{2000}, \emph{62}, 16826--16834\relax
\mciteBstWouldAddEndPuncttrue
\mciteSetBstMidEndSepPunct{\mcitedefaultmidpunct}
{\mcitedefaultendpunct}{\mcitedefaultseppunct}\relax
\EndOfBibitem
\bibitem[Corfdir \latin{et~al.}(2009)Corfdir, Lefebvre, Risti\'c, Valvin,
  Calleja, Trampert, Gani{\`{e}}re, and Deveaud-Pl\'edran]{Corfdir2009}
Corfdir,~P.; Lefebvre,~P.; Risti\'c,~J.; Valvin,~P.; Calleja,~E.; Trampert,~A.;
  Gani{\`{e}}re,~J.-D.; Deveaud-Pl\'edran,~B. {Time-resolved spectroscopy on
  {GaN} nanocolumns grown by plasma assisted molecular beam epitaxy on Si
  substrates}. \emph{J. Appl. Phys.} \textbf{2009}, \emph{105}, 013113\relax
\mciteBstWouldAddEndPuncttrue
\mciteSetBstMidEndSepPunct{\mcitedefaultmidpunct}
{\mcitedefaultendpunct}{\mcitedefaultseppunct}\relax
\EndOfBibitem
\bibitem[Hauswald \latin{et~al.}(2014)Hauswald, Corfdir, Zettler, Kaganer,
  Sabelfeld, Fern\'andez-Garrido, Flissikowski, Consonni, Gotschke, Grahn,
  Geelhaar, and Brandt]{Hauswald2014a}
Hauswald,~C.; Corfdir,~P.; Zettler,~J.~K.; Kaganer,~V.~M.; Sabelfeld,~K.~K.;
  Fern\'andez-Garrido,~S.; Flissikowski,~T.; Consonni,~V.; Gotschke,~T.;
  Grahn,~H.~T.; Geelhaar,~L.; Brandt,~O. {Origin of the nonradiative decay of
  bound excitons in {GaN} nanowires}. \emph{Phys. Rev. B} \textbf{2014},
  \emph{90}, 165304\relax
\mciteBstWouldAddEndPuncttrue
\mciteSetBstMidEndSepPunct{\mcitedefaultmidpunct}
{\mcitedefaultendpunct}{\mcitedefaultseppunct}\relax
\EndOfBibitem
\bibitem[W\"olz \latin{et~al.}(2015)W\"olz, Hauswald, Flissikowski, Gotschke,
  Fern\'andez-Garrido, Brandt, Grahn, Geelhaar, and Riechert]{Wolz2015}
W\"olz,~M.; Hauswald,~C.; Flissikowski,~T.; Gotschke,~T.;
  Fern\'andez-Garrido,~S.; Brandt,~O.; Grahn,~H.~T.; Geelhaar,~L.; Riechert,~H.
  {Epitaxial Growth of {GaN} Nanowires with High Structural Perfection on a
  Metallic TiN Film}. \emph{Nano Lett.} \textbf{2015}, \emph{15},
  3743--3747\relax
\mciteBstWouldAddEndPuncttrue
\mciteSetBstMidEndSepPunct{\mcitedefaultmidpunct}
{\mcitedefaultendpunct}{\mcitedefaultseppunct}\relax
\EndOfBibitem
\bibitem[Monemar \latin{et~al.}(2010)Monemar, Paskov, Bergman, Pozina, Toropov,
  Shubina, Malinauskas, and Usui]{Monemar2010}
Monemar,~B.; Paskov,~P.~P.; Bergman,~J.~P.; Pozina,~G.; Toropov,~A.~A.;
  Shubina,~T.~V.; Malinauskas,~T.; Usui,~A. {Transient photoluminescence of
  shallow donor bound excitons in GaN}. \emph{Phys. Rev. B} \textbf{2010},
  \emph{82}, 235202\relax
\mciteBstWouldAddEndPuncttrue
\mciteSetBstMidEndSepPunct{\mcitedefaultmidpunct}
{\mcitedefaultendpunct}{\mcitedefaultseppunct}\relax
\EndOfBibitem
\bibitem[Gorgis \latin{et~al.}(2012)Gorgis, Flissikowski, Brandt, Ch\`eze,
  Geelhaar, Riechert, and Grahn]{gorgis2012}
Gorgis,~A.; Flissikowski,~T.; Brandt,~O.; Ch\`eze,~C.; Geelhaar,~L.;
  Riechert,~H.; Grahn,~H.~T. {Time-resolved photoluminescence spectroscopy of
  individual {GaN} nanowires}. \emph{Phys. Rev. B} \textbf{2012}, \emph{86},
  041302\relax
\mciteBstWouldAddEndPuncttrue
\mciteSetBstMidEndSepPunct{\mcitedefaultmidpunct}
{\mcitedefaultendpunct}{\mcitedefaultseppunct}\relax
\EndOfBibitem
\bibitem[Consonni \latin{et~al.}(2009)Consonni, Knelangen, Jahn, Trampert,
  Geelhaar, and Riechert]{Consonni2009}
Consonni,~V.; Knelangen,~M.; Jahn,~U.; Trampert,~A.; Geelhaar,~L.; Riechert,~H.
  {Effects of nanowire coalescence on their structural and optical properties
  on a local scale}. \emph{Appl. Phys. Lett.} \textbf{2009}, \emph{95},
  241910\relax
\mciteBstWouldAddEndPuncttrue
\mciteSetBstMidEndSepPunct{\mcitedefaultmidpunct}
{\mcitedefaultendpunct}{\mcitedefaultseppunct}\relax
\EndOfBibitem
\bibitem[Grossklaus \latin{et~al.}(2013)Grossklaus, Banerjee, Jahangir,
  Bhattacharya, and Millunchick]{Grossklaus2013}
Grossklaus,~K.; Banerjee,~A.; Jahangir,~S.; Bhattacharya,~P.; Millunchick,~J.
  {Misorientation defects in coalesced self-catalyzed {GaN} nanowires}.
  \emph{J. Cryst. Growth} \textbf{2013}, \emph{371}, 142--147\relax
\mciteBstWouldAddEndPuncttrue
\mciteSetBstMidEndSepPunct{\mcitedefaultmidpunct}
{\mcitedefaultendpunct}{\mcitedefaultseppunct}\relax
\EndOfBibitem
\bibitem[Fan \latin{et~al.}(2014)Fan, Zhao, Liu, and Mi]{Fan2014}
Fan,~S.; Zhao,~S.; Liu,~X.; Mi,~Z. Study on the coalescence of dislocation-free
  GaN nanowires on Si and SiOx. \emph{Journal of Vacuum Science \& Technology
  B} \textbf{2014}, \emph{32}, 02C114\relax
\mciteBstWouldAddEndPuncttrue
\mciteSetBstMidEndSepPunct{\mcitedefaultmidpunct}
{\mcitedefaultendpunct}{\mcitedefaultseppunct}\relax
\EndOfBibitem
\bibitem[Kaganer \latin{et~al.}(2016)Kaganer, Jenichen, and
  Brandt]{Kaganer2016b}
Kaganer,~V.~M.; Jenichen,~B.; Brandt,~O. {Elastic versus Plastic Strain
  Relaxation in Coalesced {GaN} Nanowires: An X-Ray Diffraction Study}.
  \emph{Phys. Rev. Applied} \textbf{2016}, \emph{6}, 064023\relax
\mciteBstWouldAddEndPuncttrue
\mciteSetBstMidEndSepPunct{\mcitedefaultmidpunct}
{\mcitedefaultendpunct}{\mcitedefaultseppunct}\relax
\EndOfBibitem
\bibitem[Zettler \latin{et~al.}(2016)Zettler, Corfdir, Hauswald, Luna, Jahn,
  Flissikowski, Schmidt, Ronning, Trampert, Geelhaar, Grahn, Brandt, and
  Fern\'andez-Garrido]{Zettler2016}
Zettler,~J.~K.; Corfdir,~P.; Hauswald,~C.; Luna,~E.; Jahn,~U.;
  Flissikowski,~T.; Schmidt,~E.; Ronning,~C.; Trampert,~A.; Geelhaar,~L.;
  Grahn,~H.~T.; Brandt,~O.; Fern\'andez-Garrido,~S. {Observation of
  Dielectrically Confined Excitons in Ultrathin {GaN} Nanowires up to Room
  Temperature}. \emph{Nano Lett.} \textbf{2016}, \emph{16}, 973--980\relax
\mciteBstWouldAddEndPuncttrue
\mciteSetBstMidEndSepPunct{\mcitedefaultmidpunct}
{\mcitedefaultendpunct}{\mcitedefaultseppunct}\relax
\EndOfBibitem
\bibitem[Burton \latin{et~al.}(1951)Burton, Cabrera, Frank, and
  Mott]{Burton1951}
Burton,~W.~K.; Cabrera,~N.; Frank,~F.~C.; Mott,~N.~F. {The growth of crystals
  and the equilibrium structure of their surfaces}. \emph{Phil. Trans. R. Soc.
  A} \textbf{1951}, \emph{243}, 299--358\relax
\mciteBstWouldAddEndPuncttrue
\mciteSetBstMidEndSepPunct{\mcitedefaultmidpunct}
{\mcitedefaultendpunct}{\mcitedefaultseppunct}\relax
\EndOfBibitem
\bibitem[Ishizaka and Murata(1994)Ishizaka, and Murata]{Ishizaka1994}
Ishizaka,~A.; Murata,~Y. {Crystal growth model for molecular beam epitaxy: Role
  of kinks on crystal growth}. \emph{J. Phys.: Condens. Matter} \textbf{1994},
  \emph{6}, L693--L698\relax
\mciteBstWouldAddEndPuncttrue
\mciteSetBstMidEndSepPunct{\mcitedefaultmidpunct}
{\mcitedefaultendpunct}{\mcitedefaultseppunct}\relax
\EndOfBibitem
\bibitem[Zettler \latin{et~al.}(2015)Zettler, Hauswald, Corfdir, Musolino,
  Geelhaar, Riechert, Brandt, and Fern{\'{a}}ndez-Garrido]{Zettler2015b}
Zettler,~J.~K.; Hauswald,~C.; Corfdir,~P.; Musolino,~M.; Geelhaar,~L.;
  Riechert,~H.; Brandt,~O.; Fern{\'{a}}ndez-Garrido,~S. {High-Temperature
  Growth of {GaN} Nanowires by Molecular Beam Epitaxy: Toward the Material
  Quality of Bulk GaN}. \emph{Cryst. Growth Des.} \textbf{2015}, \emph{15},
  4104--4109\relax
\mciteBstWouldAddEndPuncttrue
\mciteSetBstMidEndSepPunct{\mcitedefaultmidpunct}
{\mcitedefaultendpunct}{\mcitedefaultseppunct}\relax
\EndOfBibitem
\bibitem[Calabrese \latin{et~al.}(2019)Calabrese, Gao, van Treeck, Corfdir,
  Sinito, Auzelle, Trampert, Geelhaar, Brandt, and
  Fern{\'{a}}ndez-Garrido]{Calabrese2019}
Calabrese,~G.; Gao,~G.; van Treeck,~D.; Corfdir,~P.; Sinito,~C.; Auzelle,~T.;
  Trampert,~A.; Geelhaar,~L.; Brandt,~O.; Fern{\'{a}}ndez-Garrido,~S.
  Interfacial reactions during the molecular beam epitaxy of {GaN} nanowires on
  {Ti/Al2O3}. \emph{Nanotechnology} \textbf{2019}, \emph{30}, 114001\relax
\mciteBstWouldAddEndPuncttrue
\mciteSetBstMidEndSepPunct{\mcitedefaultmidpunct}
{\mcitedefaultendpunct}{\mcitedefaultseppunct}\relax
\EndOfBibitem
\bibitem[Sarwar \latin{et~al.}(2015)Sarwar, Carnevale, Yang, Kent, Jamison,
  McComb, and Myers]{Sarwar2015}
Sarwar,~A.~G.; Carnevale,~S.~D.; Yang,~F.; Kent,~T.~F.; Jamison,~J.~J.;
  McComb,~D.~W.; Myers,~R.~C. {Semiconductor Nanowire Light-Emitting Diodes
  Grown on Metal: A Direction Toward Large-Scale Fabrication of Nanowire
  Devices}. \emph{Small} \textbf{2015}, \emph{11}, 5402--5408\relax
\mciteBstWouldAddEndPuncttrue
\mciteSetBstMidEndSepPunct{\mcitedefaultmidpunct}
{\mcitedefaultendpunct}{\mcitedefaultseppunct}\relax
\EndOfBibitem
\bibitem[Zhao \latin{et~al.}(2016)Zhao, Ng, Wei, Prabaswara, Alias, Janjua,
  Shen, and Ooi]{Zhao2016}
Zhao,~C.; Ng,~T.~K.; Wei,~N.; Prabaswara,~A.; Alias,~M.~S.; Janjua,~B.;
  Shen,~C.; Ooi,~B.~S. {Facile Formation of High-Quality {InGaN}/{GaN}
  Quantum-Disks-in-Nanowires on Bulk-Metal Substrates for High-Power
  Light-Emitters}. \emph{Nano Lett.} \textbf{2016}, \emph{16}, 1056--1063\relax
\mciteBstWouldAddEndPuncttrue
\mciteSetBstMidEndSepPunct{\mcitedefaultmidpunct}
{\mcitedefaultendpunct}{\mcitedefaultseppunct}\relax
\EndOfBibitem
\bibitem[Calabrese \latin{et~al.}(2016)Calabrese, Corfdir, Gao, Pf\"{u}ller,
  Trampert, Brandt, Geelhaar, and Fern\'andez-Garrido]{Calabrese2016}
Calabrese,~G.; Corfdir,~P.; Gao,~G.; Pf\"{u}ller,~C.; Trampert,~A.; Brandt,~O.;
  Geelhaar,~L.; Fern\'andez-Garrido,~S. {Molecular beam epitaxy of single
  crystalline {GaN} nanowires on a flexible Ti foil}. \emph{Appl. Phys. Lett.}
  \textbf{2016}, \emph{108}, 202101\relax
\mciteBstWouldAddEndPuncttrue
\mciteSetBstMidEndSepPunct{\mcitedefaultmidpunct}
{\mcitedefaultendpunct}{\mcitedefaultseppunct}\relax
\EndOfBibitem
\bibitem[Ramesh \latin{et~al.}(2019)Ramesh, Tyagi, Abhiram, Gupta, Kumar, and
  Kushvaha]{Ramesh2019}
Ramesh,~C.; Tyagi,~P.; Abhiram,~G.; Gupta,~G.; Kumar,~M.~S.; Kushvaha,~S. {Role
  of growth temperature on formation of single crystalline {GaN} nanorods on
  flexible titanium foil by laser molecular beam epitaxy}. \emph{J. Cryst.
  Growth} \textbf{2019}, \emph{509}, 23--28\relax
\mciteBstWouldAddEndPuncttrue
\mciteSetBstMidEndSepPunct{\mcitedefaultmidpunct}
{\mcitedefaultendpunct}{\mcitedefaultseppunct}\relax
\EndOfBibitem
\bibitem[Choi \latin{et~al.}(2011)Choi, Zoulkarneev, Kim, Baik, Yang, Park,
  Suh, Kim, Bin~Son, Lee, Kim, Kim, and Kim]{Choi2011}
Choi,~J.~H.; Zoulkarneev,~A.; Kim,~S.~I.; Baik,~C.~W.; Yang,~M.~H.;
  Park,~S.~S.; Suh,~H.; Kim,~U.~J.; Bin~Son,~H.; Lee,~J.~S.; Kim,~M.;
  Kim,~J.~M.; Kim,~K. Nearly Single-Crystalline {{GaN}} Light-Emitting Diodes
  on Amorphous Glass Substrates. \emph{Nat. Photonics} \textbf{2011}, \emph{5},
  763--769\relax
\mciteBstWouldAddEndPuncttrue
\mciteSetBstMidEndSepPunct{\mcitedefaultmidpunct}
{\mcitedefaultendpunct}{\mcitedefaultseppunct}\relax
\EndOfBibitem
\bibitem[Naik \latin{et~al.}(2011)Naik, Kim, and Boltasseva]{Naik2011}
Naik,~G.~V.; Kim,~J.; Boltasseva,~A. Oxides and Nitrides as Alternative
  Plasmonic Materials in the Optical Range [{{Invited}}]. \emph{Opt. Mater.
  Express} \textbf{2011}, \emph{1}, 1090--1099\relax
\mciteBstWouldAddEndPuncttrue
\mciteSetBstMidEndSepPunct{\mcitedefaultmidpunct}
{\mcitedefaultendpunct}{\mcitedefaultseppunct}\relax
\EndOfBibitem
\bibitem[Li \latin{et~al.}(2014)Li, Guler, Kinsey, Naik, Boltasseva, Guan,
  Shalaev, and Kildishev]{Li2014}
Li,~W.; Guler,~U.; Kinsey,~N.; Naik,~G.~V.; Boltasseva,~A.; Guan,~J.;
  Shalaev,~V.~M.; Kildishev,~A.~V. Refractory {{Plasmonics}} with {{Titanium
  Nitride}}: {{Broadband Metamaterial Absorber}}. \emph{Adv. Mater.}
  \textbf{2014}, \emph{26}, 7959--7965\relax
\mciteBstWouldAddEndPuncttrue
\mciteSetBstMidEndSepPunct{\mcitedefaultmidpunct}
{\mcitedefaultendpunct}{\mcitedefaultseppunct}\relax
\EndOfBibitem
\bibitem[May \latin{et~al.}(2016)May, Sarwar, and Myers]{May2016}
May,~B.~J.; Sarwar,~A. T. M.~G.; Myers,~R.~C. Nanowire {{LEDs}} Grown Directly
  on Flexible Metal Foil. \emph{Appl. Phys. Lett.} \textbf{2016}, \emph{108},
  141103\relax
\mciteBstWouldAddEndPuncttrue
\mciteSetBstMidEndSepPunct{\mcitedefaultmidpunct}
{\mcitedefaultendpunct}{\mcitedefaultseppunct}\relax
\EndOfBibitem
\bibitem[Lu \latin{et~al.}(2017)Lu, Sokhoyan, Cheng, Kafaie~Shirmanesh,
  Davoyan, Pala, Thyagarajan, and Atwater]{Lu2017}
Lu,~Y.-J.; Sokhoyan,~R.; Cheng,~W.-H.; Kafaie~Shirmanesh,~G.; Davoyan,~A.~R.;
  Pala,~R.~A.; Thyagarajan,~K.; Atwater,~H.~A. Dynamically Controlled
  {{Purcell}} Enhancement of Visible Spontaneous Emission in a Gated Plasmonic
  Heterostructure. \emph{Nat. Commun.} \textbf{2017}, \emph{8}, 1631\relax
\mciteBstWouldAddEndPuncttrue
\mciteSetBstMidEndSepPunct{\mcitedefaultmidpunct}
{\mcitedefaultendpunct}{\mcitedefaultseppunct}\relax
\EndOfBibitem
\bibitem[Fern{\'{a}}ndez-Garrido \latin{et~al.}(2015)Fern{\'{a}}ndez-Garrido,
  Zettler, Geelhaar, and Brandt]{Garrido2015}
Fern{\'{a}}ndez-Garrido,~S.; Zettler,~J.~K.; Geelhaar,~L.; Brandt,~O.
  {Monitoring the Formation of Nanowires by Line-of-Sight Quadrupole Mass
  Spectrometry: A Comprehensive Description of the Temporal Evolution of {GaN}
  Nanowire Ensembles}. \emph{Nano Lett.} \textbf{2015}, \emph{15},
  1930--1937\relax
\mciteBstWouldAddEndPuncttrue
\mciteSetBstMidEndSepPunct{\mcitedefaultmidpunct}
{\mcitedefaultendpunct}{\mcitedefaultseppunct}\relax
\EndOfBibitem
\bibitem[Zettler \latin{et~al.}(2015)Zettler, Corfdir, Geelhaar, Riechert,
  Brandt, and Fern{\'{a}}ndez-Garrido]{Zettler2015a}
Zettler,~J.~K.; Corfdir,~P.; Geelhaar,~L.; Riechert,~H.; Brandt,~O.;
  Fern{\'{a}}ndez-Garrido,~S. Improved control over spontaneously formed {GaN}
  nanowires in molecular beam epitaxy using a two-step growth process.
  \emph{Nanotechnology} \textbf{2015}, \emph{26}, 445604\relax
\mciteBstWouldAddEndPuncttrue
\mciteSetBstMidEndSepPunct{\mcitedefaultmidpunct}
{\mcitedefaultendpunct}{\mcitedefaultseppunct}\relax
\EndOfBibitem
\bibitem[Spengler \latin{et~al.}(1978)Spengler, Kaiser, Christensen, and
  M\"uller-Vogt]{Spengler1978}
Spengler,~W.; Kaiser,~R.; Christensen,~A.~N.; M\"uller-Vogt,~G. {Raman
  scattering, superconductivity, and phonon density of states of stoichiometric
  and nonstoichiometric TiN}. \emph{Phys. Rev. B} \textbf{1978}, \emph{17},
  1095--1101\relax
\mciteBstWouldAddEndPuncttrue
\mciteSetBstMidEndSepPunct{\mcitedefaultmidpunct}
{\mcitedefaultendpunct}{\mcitedefaultseppunct}\relax
\EndOfBibitem
\bibitem[Brandt \latin{et~al.}(2014)Brandt, Fern{\'{a}}ndez-Garrido, Zettler,
  Luna, Jahn, Ch{\`{e}}ze, and Kaganer]{Brandt2014}
Brandt,~O.; Fern{\'{a}}ndez-Garrido,~S.; Zettler,~J.~K.; Luna,~E.; Jahn,~U.;
  Ch{\`{e}}ze,~C.; Kaganer,~V.~M. {Statistical Analysis of the Shape of
  One-Dimensional Nanostructures: Determining the Coalescence Degree of
  Spontaneously Formed {GaN} Nanowires}. \emph{Cryst. Growth Des.}
  \textbf{2014}, \emph{14}, 2246--2253\relax
\mciteBstWouldAddEndPuncttrue
\mciteSetBstMidEndSepPunct{\mcitedefaultmidpunct}
{\mcitedefaultendpunct}{\mcitedefaultseppunct}\relax
\EndOfBibitem
\bibitem[van Treeck \latin{et~al.}(2018)van Treeck, Calabrese, Goertz, Kaganer,
  Brandt, Fern{\'a}ndez-Garrido, and Geelhaar]{vanTreeck2018}
van Treeck,~D.; Calabrese,~G.; Goertz,~J. J.~W.; Kaganer,~V.~M.; Brandt,~O.;
  Fern{\'a}ndez-Garrido,~S.; Geelhaar,~L. {Self-assembled formation of long,
  thin, and uncoalesced {GaN} nanowires on crystalline TiN films}. \emph{Nano
  Res.} \textbf{2018}, \emph{11}, 565--576\relax
\mciteBstWouldAddEndPuncttrue
\mciteSetBstMidEndSepPunct{\mcitedefaultmidpunct}
{\mcitedefaultendpunct}{\mcitedefaultseppunct}\relax
\EndOfBibitem
\bibitem[Kaganer \latin{et~al.}(2016)Kaganer, Fern\'andez-Garrido, Dogan,
  Sabelfeld, and Brandt]{Kaganer2016a}
Kaganer,~V.~M.; Fern\'andez-Garrido,~S.; Dogan,~P.; Sabelfeld,~K.~K.;
  Brandt,~O. {Nucleation, Growth, and Bundling of {GaN} Nanowires in Molecular
  Beam Epitaxy: Disentangling the Origin of Nanowire Coalescence}. \emph{Nano
  Lett.} \textbf{2016}, \emph{16}, 3717--3725\relax
\mciteBstWouldAddEndPuncttrue
\mciteSetBstMidEndSepPunct{\mcitedefaultmidpunct}
{\mcitedefaultendpunct}{\mcitedefaultseppunct}\relax
\EndOfBibitem
\bibitem[Davydov \latin{et~al.}(1997)Davydov, Averkiev, Goncharuk, Nelson,
  Nikitina, Polkovnikov, Smirnov, Jacobson, and Semchinova]{Davydov1997}
Davydov,~V.~Y.; Averkiev,~N.~S.; Goncharuk,~I.~N.; Nelson,~D.~K.;
  Nikitina,~I.~P.; Polkovnikov,~A.~S.; Smirnov,~A.~N.; Jacobson,~M.~A.;
  Semchinova,~O.~K. Raman and photoluminescence studies of biaxial strain in
  GaN epitaxial layers grown on 6H–SiC. \emph{Journal of Applied Physics}
  \textbf{1997}, \emph{82}\relax
\mciteBstWouldAddEndPuncttrue
\mciteSetBstMidEndSepPunct{\mcitedefaultmidpunct}
{\mcitedefaultendpunct}{\mcitedefaultseppunct}\relax
\EndOfBibitem
\bibitem[Azadmand \latin{et~al.}(2020)Azadmand, Auzelle, L{\"{a}}hnemann, Gao,
  Nicolai, Ramsteiner, Trampert, Sanguinetti, Brandt, and
  Geelhaar]{Azadmand2020}
Azadmand,~M.; Auzelle,~T.; L{\"{a}}hnemann,~J.; Gao,~G.; Nicolai,~L.;
  Ramsteiner,~M.; Trampert,~A.; Sanguinetti,~S.; Brandt,~O.; Geelhaar,~L.
  {Self-assembly of well-separated {AlN} nanowires directly on sputtered
  metallic TiN films}. \emph{Phys. Status Solidi Rapid Res. Lett.}
  \textbf{2020}, \emph{14}, 1900615\relax
\mciteBstWouldAddEndPuncttrue
\mciteSetBstMidEndSepPunct{\mcitedefaultmidpunct}
{\mcitedefaultendpunct}{\mcitedefaultseppunct}\relax
\EndOfBibitem
\bibitem[P\'ecz \latin{et~al.}(1995)P\'ecz, Frangis, Logothetidis, Alexandrou,
  Barna, and Stoemenos]{Pecz1995}
P\'ecz,~B.; Frangis,~N.; Logothetidis,~S.; Alexandrou,~I.; Barna,~P.;
  Stoemenos,~J. Electron microscopy characterization of {TiN} films on {Si},
  grown by dc reactive magnetron sputtering. \emph{Thin Solid Films}
  \textbf{1995}, \emph{268}, 57--63\relax
\mciteBstWouldAddEndPuncttrue
\mciteSetBstMidEndSepPunct{\mcitedefaultmidpunct}
{\mcitedefaultendpunct}{\mcitedefaultseppunct}\relax
\EndOfBibitem
\bibitem[Brandt \latin{et~al.}(2010)Brandt, Pf\"uller, Ch\`eze, Geelhaar, and
  Riechert]{Brandt2010}
Brandt,~O.; Pf\"uller,~C.; Ch\`eze,~C.; Geelhaar,~L.; Riechert,~H. {Sub-meV
  linewidth of excitonic luminescence in single {GaN} nanowires: Direct
  evidence for surface excitons}. \emph{Phys. Rev. B} \textbf{2010}, \emph{81},
  045302\relax
\mciteBstWouldAddEndPuncttrue
\mciteSetBstMidEndSepPunct{\mcitedefaultmidpunct}
{\mcitedefaultendpunct}{\mcitedefaultseppunct}\relax
\EndOfBibitem
\bibitem[Corfdir \latin{et~al.}(2014)Corfdir, Zettler, Hauswald,
  Fern\'andez-Garrido, Brandt, and Lefebvre]{Corfdir2014a}
Corfdir,~P.; Zettler,~J.~K.; Hauswald,~C.; Fern\'andez-Garrido,~S.; Brandt,~O.;
  Lefebvre,~P. {Sub-meV linewidth in {GaN} nanowire ensembles: Absence of
  surface excitons due to the field ionization of donors}. \emph{Phys. Rev. B}
  \textbf{2014}, \emph{90}, 205301\relax
\mciteBstWouldAddEndPuncttrue
\mciteSetBstMidEndSepPunct{\mcitedefaultmidpunct}
{\mcitedefaultendpunct}{\mcitedefaultseppunct}\relax
\EndOfBibitem
\bibitem[Calabrese \latin{et~al.}(2020)Calabrese, van Treeck, Kaganer,
  Konovalov, Corfdir, Sinito, Geelhaar, Brandt, and
  Fern{\'{a}}ndez-Garrido]{Calabrese2020}
Calabrese,~G.; van Treeck,~D.; Kaganer,~V.; Konovalov,~O.; Corfdir,~P.;
  Sinito,~C.; Geelhaar,~L.; Brandt,~O.; Fern{\'{a}}ndez-Garrido,~S.
  Radius-dependent homogeneous strain in uncoalesced {GaN} nanowires.
  \emph{Acta Mater.} \textbf{2020}, \emph{195}, 87--97\relax
\mciteBstWouldAddEndPuncttrue
\mciteSetBstMidEndSepPunct{\mcitedefaultmidpunct}
{\mcitedefaultendpunct}{\mcitedefaultseppunct}\relax
\EndOfBibitem
\bibitem[Pf{\"{u}}ller \latin{et~al.}(2010)Pf{\"{u}}ller, Brandt, Flissikowski,
  Ch{\`{e}}ze, Geelhaar, Grahn, and Riechert]{Pfuller2010a}
Pf{\"{u}}ller,~C.; Brandt,~O.; Flissikowski,~T.; Ch{\`{e}}ze,~C.; Geelhaar,~L.;
  Grahn,~H.~T.; Riechert,~H. {Statistical analysis of excitonic transitions in
  single, free-standing {GaN} nanowires: Probing impurity incorporation in the
  Poissonian limit}. \emph{Nano Res.} \textbf{2010}, \emph{3}, 881--888\relax
\mciteBstWouldAddEndPuncttrue
\mciteSetBstMidEndSepPunct{\mcitedefaultmidpunct}
{\mcitedefaultendpunct}{\mcitedefaultseppunct}\relax
\EndOfBibitem
\bibitem[Hauswald \latin{et~al.}(2013)Hauswald, Flissikowski, Gotschke,
  Calarco, Geelhaar, Grahn, and Brandt]{Hauswald2013}
Hauswald,~C.; Flissikowski,~T.; Gotschke,~T.; Calarco,~R.; Geelhaar,~L.;
  Grahn,~H.~T.; Brandt,~O. {Coupling of exciton states as the origin of their
  biexponential decay dynamics in {GaN} nanowires}. \emph{Phys. Rev. B}
  \textbf{2013}, \emph{88}, 075312\relax
\mciteBstWouldAddEndPuncttrue
\mciteSetBstMidEndSepPunct{\mcitedefaultmidpunct}
{\mcitedefaultendpunct}{\mcitedefaultseppunct}\relax
\EndOfBibitem
\bibitem[Calarco \latin{et~al.}(2005)Calarco, Marso, Richter, Aykanat, Meijers,
  {V D Hart}, Stoica, and L{\"{u}}th]{Calarco2005}
Calarco,~R.; Marso,~M.; Richter,~T.; Aykanat,~A.~I.; Meijers,~R.~J.; {V D
  Hart},~A.; Stoica,~T.; L{\"{u}}th,~H. {Size-dependent photoconductivity in
  {MBE}-grown {GaN}-nanowires.} \emph{Nano Lett.} \textbf{2005}, \emph{5},
  981--984\relax
\mciteBstWouldAddEndPuncttrue
\mciteSetBstMidEndSepPunct{\mcitedefaultmidpunct}
{\mcitedefaultendpunct}{\mcitedefaultseppunct}\relax
\EndOfBibitem
\bibitem[Portz \latin{et~al.}(2018)Portz, Schnedler, Eisele, Dunin-Borkowski,
  and Ebert]{Portz2018}
Portz,~V.; Schnedler,~M.; Eisele,~H.; Dunin-Borkowski,~R.~E.; Ebert,~P.
  {Electron affinity and surface states of {GaN} $m$-plane facets: Implication
  for electronic self-passivation}. \emph{Phys. Rev. B} \textbf{2018},
  \emph{97}, 115433\relax
\mciteBstWouldAddEndPuncttrue
\mciteSetBstMidEndSepPunct{\mcitedefaultmidpunct}
{\mcitedefaultendpunct}{\mcitedefaultseppunct}\relax
\EndOfBibitem
\bibitem[Pf{\"{u}}ller \latin{et~al.}(2010)Pf{\"{u}}ller, Brandt, Grosse,
  Flissikowski, Ch{\`{e}}ze, Consonni, Geelhaar, Grahn, and
  Riechert]{Pfuller2010b}
Pf{\"{u}}ller,~C.; Brandt,~O.; Grosse,~F.; Flissikowski,~T.; Ch{\`{e}}ze,~C.;
  Consonni,~V.; Geelhaar,~L.; Grahn,~H.~T.; Riechert,~H. {Unpinning the Fermi
  level of {GaN} nanowires by ultraviolet radiation}. \emph{Phys. Rev. B}
  \textbf{2010}, \emph{82}, 045320\relax
\mciteBstWouldAddEndPuncttrue
\mciteSetBstMidEndSepPunct{\mcitedefaultmidpunct}
{\mcitedefaultendpunct}{\mcitedefaultseppunct}\relax
\EndOfBibitem
\bibitem[Corfdir \latin{et~al.}(2014)Corfdir, Hauswald, Zettler, Flissikowski,
  L{\"{a}}hnemann, Fern{\'{a}}ndez-Garrido, Geelhaar, Grahn, and
  Brandt]{Corfdir2014b}
Corfdir,~P.; Hauswald,~C.; Zettler,~J.~K.; Flissikowski,~T.;
  L{\"{a}}hnemann,~J.; Fern{\'{a}}ndez-Garrido,~S.; Geelhaar,~L.; Grahn,~H.~T.;
  Brandt,~O. {Stacking faults as quantum wells in nanowires: Density of states,
  oscillator strength and radiative efficiency}. \emph{Phys. Rev. B}
  \textbf{2014}, \emph{90}, 195309\relax
\mciteBstWouldAddEndPuncttrue
\mciteSetBstMidEndSepPunct{\mcitedefaultmidpunct}
{\mcitedefaultendpunct}{\mcitedefaultseppunct}\relax
\EndOfBibitem
\bibitem[Blossey(1970)]{Blossey1970}
Blossey,~D.~F. {Wannier Exciton in an Electric Field. I. Optical Absorption by
  Bound and Continuum States}. \emph{Phys. Rev. B} \textbf{1970}, \emph{2},
  3976--3990\relax
\mciteBstWouldAddEndPuncttrue
\mciteSetBstMidEndSepPunct{\mcitedefaultmidpunct}
{\mcitedefaultendpunct}{\mcitedefaultseppunct}\relax
\EndOfBibitem
\bibitem[Winnerl \latin{et~al.}(2015)Winnerl, Pereira, and
  Stutzmann]{Winnerl2015}
Winnerl,~A.; Pereira,~R.~N.; Stutzmann,~M. Kinetics of optically excited charge
  carriers at the GaN surface. \emph{Phys. Rev. B} \textbf{2015}, \emph{91},
  075316\relax
\mciteBstWouldAddEndPuncttrue
\mciteSetBstMidEndSepPunct{\mcitedefaultmidpunct}
{\mcitedefaultendpunct}{\mcitedefaultseppunct}\relax
\EndOfBibitem
\bibitem[Yamabe \latin{et~al.}(1977)Yamabe, Tachibana, and
  Silverstone]{Yamabe1977}
Yamabe,~T.; Tachibana,~A.; Silverstone,~H.~J. Theory of the ionization of the
  hydrogen atom by an external electrostatic field. \emph{Phys. Rev. A}
  \textbf{1977}, \emph{16}, 877--890\relax
\mciteBstWouldAddEndPuncttrue
\mciteSetBstMidEndSepPunct{\mcitedefaultmidpunct}
{\mcitedefaultendpunct}{\mcitedefaultseppunct}\relax
\EndOfBibitem
\bibitem[Banavar \latin{et~al.}(1979)Banavar, Coon, and Derkits]{Banavar1979}
Banavar,~J.~R.; Coon,~D.~D.; Derkits,~G.~E. Low‐temperature field ionization
  of localized impurity levels in semiconductors. \emph{Appl. Phys. Lett.}
  \textbf{1979}, \emph{34}, 94--96\relax
\mciteBstWouldAddEndPuncttrue
\mciteSetBstMidEndSepPunct{\mcitedefaultmidpunct}
{\mcitedefaultendpunct}{\mcitedefaultseppunct}\relax
\EndOfBibitem
\bibitem[Kaganer \latin{et~al.}(2018)Kaganer, Sabelfeld, and
  Brandt]{Kaganer2018}
Kaganer,~V.~M.; Sabelfeld,~K.~K.; Brandt,~O. {Piezoelectric field, exciton
  lifetime, and cathodoluminescence intensity at threading dislocations in
  {GaN}\{0001\}}. \emph{Appl. Phys. Lett.} \textbf{2018}, \emph{112},
  122101\relax
\mciteBstWouldAddEndPuncttrue
\mciteSetBstMidEndSepPunct{\mcitedefaultmidpunct}
{\mcitedefaultendpunct}{\mcitedefaultseppunct}\relax
\EndOfBibitem
\bibitem[Schlager \latin{et~al.}(2008)Schlager, Bertness, Blanchard, Robins,
  Roshko, and Sanford]{Schlager2008}
Schlager,~J.~B.; Bertness,~K.~A.; Blanchard,~P.~T.; Robins,~L.~H.; Roshko,~A.;
  Sanford,~N.~A. {Steady-state and time-resolved photoluminescence from relaxed
  and strained {GaN} nanowires grown by catalyst-free molecular-beam epitaxy}.
  \emph{J. Appl. Phys.} \textbf{2008}, \emph{103}, 124309\relax
\mciteBstWouldAddEndPuncttrue
\mciteSetBstMidEndSepPunct{\mcitedefaultmidpunct}
{\mcitedefaultendpunct}{\mcitedefaultseppunct}\relax
\EndOfBibitem
\bibitem[Schlager \latin{et~al.}(2011)Schlager, Sanford, Bertness, and
  Roshko]{Schlager2011}
Schlager,~J.~B.; Sanford,~N.~A.; Bertness,~K.~A.; Roshko,~A.
  {Injection-level-dependent internal quantum efficiency and lasing in
  low-defect {GaN} nanowires}. \emph{J. Appl. Phys.} \textbf{2011}, \emph{109},
  044312\relax
\mciteBstWouldAddEndPuncttrue
\mciteSetBstMidEndSepPunct{\mcitedefaultmidpunct}
{\mcitedefaultendpunct}{\mcitedefaultseppunct}\relax
\EndOfBibitem
\bibitem[Schultheis and Lagois(1984)Schultheis, and Lagois]{Schultheis1984}
Schultheis,~L.; Lagois,~J. Excitonic polaritons in electric fields at {GaAs}
  surfaces. \emph{Phys. Rev. B} \textbf{1984}, \emph{29}, 6784--6790\relax
\mciteBstWouldAddEndPuncttrue
\mciteSetBstMidEndSepPunct{\mcitedefaultmidpunct}
{\mcitedefaultendpunct}{\mcitedefaultseppunct}\relax
\EndOfBibitem
\bibitem[Shokhovets \latin{et~al.}(2003)Shokhovets, Fuhrmann, Goldhahn, Gobsch,
  Ambacher, Hermann, Karrer, and Eickhoff]{Shokhovets2003}
Shokhovets,~S.; Fuhrmann,~D.; Goldhahn,~R.; Gobsch,~G.; Ambacher,~O.;
  Hermann,~M.; Karrer,~U.; Eickhoff,~M. Exciton quenching in Pt/{GaN}
  {Schottky} diodes with Ga- and N-face polarity. \emph{Appl. Phys. Lett.}
  \textbf{2003}, \emph{82}, 1712--1714\relax
\mciteBstWouldAddEndPuncttrue
\mciteSetBstMidEndSepPunct{\mcitedefaultmidpunct}
{\mcitedefaultendpunct}{\mcitedefaultseppunct}\relax
\EndOfBibitem
\bibitem[Zhao \latin{et~al.}(2015)Zhao, Ng, Prabaswara, Conroy, Jahangir,
  Frost, O'Connell, Holmes, Parbrook, Bhattacharya, and Ooi]{Zhao2015}
Zhao,~C.; Ng,~T.~K.; Prabaswara,~A.; Conroy,~M.; Jahangir,~S.; Frost,~T.;
  O'Connell,~J.; Holmes,~J.~D.; Parbrook,~P.~J.; Bhattacharya,~P.; Ooi,~B.~S.
  {An enhanced surface passivation effect in InGaN/GaN disk-in-nanowire light
  emitting diodes for mitigating Shockley–Read–Hall recombination}.
  \emph{Nanoscale} \textbf{2015}, \emph{7}, 16658--16665\relax
\mciteBstWouldAddEndPuncttrue
\mciteSetBstMidEndSepPunct{\mcitedefaultmidpunct}
{\mcitedefaultendpunct}{\mcitedefaultseppunct}\relax
\EndOfBibitem
\bibitem[Varadhan \latin{et~al.}(2017)Varadhan, Fu, Priante, Retamal, Zhao,
  Ebaid, Ng, Ajia, Mitra, Roqan, Ooi, and He]{Varadhan2017}
Varadhan,~P.; Fu,~H.-C.; Priante,~D.; Retamal,~J. R.~D.; Zhao,~C.; Ebaid,~M.;
  Ng,~T.~K.; Ajia,~I.; Mitra,~S.; Roqan,~I.~S.; Ooi,~B.~S.; He,~J.-H. {Surface
  Passivation of GaN Nanowires for Enhanced Photoelectrochemical
  Water-Splitting}. \emph{Nano Lett.} \textbf{2017}, \emph{17},
  1520--1528\relax
\mciteBstWouldAddEndPuncttrue
\mciteSetBstMidEndSepPunct{\mcitedefaultmidpunct}
{\mcitedefaultendpunct}{\mcitedefaultseppunct}\relax
\EndOfBibitem
\bibitem[Hofmann and Rinke(2017)Hofmann, and Rinke]{Hofmann2017}
Hofmann,~O.~T.; Rinke,~P. {Band Bending Engineering at Organic/Inorganic
  Interfaces Using Organic Self-Assembled Monolayers}. \emph{Adv. Electron.
  Mater.} \textbf{2017}, \emph{3}, 1600373\relax
\mciteBstWouldAddEndPuncttrue
\mciteSetBstMidEndSepPunct{\mcitedefaultmidpunct}
{\mcitedefaultendpunct}{\mcitedefaultseppunct}\relax
\EndOfBibitem
\bibitem[Wang \latin{et~al.}(2017)Wang, Hao, Yu, Wu, Wang, Wang, Sun, Xiong,
  Han, Li, and Luo]{Wang2017}
Wang,~Z.~L.; Hao,~Z.~B.; Yu,~J.~D.; Wu,~C.; Wang,~L.; Wang,~J.; Sun,~C.~Z.;
  Xiong,~B.; Han,~Y.~J.; Li,~H.~T.; Luo,~Y. {Manipulating the Band Bending of
  InGaN/GaN Quantum Dots in Nanowires by Surface Passivation}. \emph{J. Phys.
  Chem. C} \textbf{2017}, \emph{121}, 6380--6385\relax
\mciteBstWouldAddEndPuncttrue
\mciteSetBstMidEndSepPunct{\mcitedefaultmidpunct}
{\mcitedefaultendpunct}{\mcitedefaultseppunct}\relax
\EndOfBibitem
\bibitem[Latzel \latin{et~al.}(2017)Latzel, B{\"{u}}ttner, Sarau,
  H{\"{o}}flich, Heilmann, Chen, Wen, Conibeer, and Christiansen]{Latzel2017}
Latzel,~M.; B{\"{u}}ttner,~P.; Sarau,~G.; H{\"{o}}flich,~K.; Heilmann,~M.;
  Chen,~W.; Wen,~X.; Conibeer,~G.; Christiansen,~S.~H. {Significant performance
  enhancement of InGaN/GaN nanorod LEDs with multi-layer graphene transparent
  electrodes by alumina surface passivation}. \emph{Nanotechnology}
  \textbf{2017}, \emph{28}, 055201\relax
\mciteBstWouldAddEndPuncttrue
\mciteSetBstMidEndSepPunct{\mcitedefaultmidpunct}
{\mcitedefaultendpunct}{\mcitedefaultseppunct}\relax
\EndOfBibitem
\bibitem[Wong \latin{et~al.}(2018)Wong, Hwang, Alhassan, Lee, Ley, Nakamura,
  and DenBaars]{Wong2018}
Wong,~M.~S.; Hwang,~D.; Alhassan,~A.~I.; Lee,~C.; Ley,~R.; Nakamura,~S.;
  DenBaars,~S.~P. {High efficiency of III-nitride micro-light-emitting diodes
  by sidewall passivation using atomic layer deposition}. \emph{Opt. Express}
  \textbf{2018}, \emph{26}, 21324--21331\relax
\mciteBstWouldAddEndPuncttrue
\mciteSetBstMidEndSepPunct{\mcitedefaultmidpunct}
{\mcitedefaultendpunct}{\mcitedefaultseppunct}\relax
\EndOfBibitem
\bibitem[Auzelle \latin{et~al.}(2021)Auzelle, Ullrich, Hietzschold, Sinito,
  Brackmann, Kowalsky, Mankel, Brandt, Lovrincic, and
  Fernández-Garrido]{Auzelle2020}
Auzelle,~T.; Ullrich,~F.; Hietzschold,~S.; Sinito,~C.; Brackmann,~S.;
  Kowalsky,~W.; Mankel,~E.; Brandt,~O.; Lovrincic,~R.; Fernández-Garrido,~S.
  External Control of GaN Band Bending Using Phosphonate Self-Assembled
  Monolayers. \emph{ACS Applied Materials \& Interfaces} \textbf{2021},
  \emph{13}, 4626--4635, PMID: 33439013\relax
\mciteBstWouldAddEndPuncttrue
\mciteSetBstMidEndSepPunct{\mcitedefaultmidpunct}
{\mcitedefaultendpunct}{\mcitedefaultseppunct}\relax
\EndOfBibitem
\bibitem[Hetzl \latin{et~al.}(2017)Hetzl, Winnerl, Francaviglia, Kraut,
  Döblinger, Matich, Fontcuberta~i Morral, and Stutzmann]{Hetzl2017}
Hetzl,~M.; Winnerl,~J.; Francaviglia,~L.; Kraut,~M.; Döblinger,~M.;
  Matich,~S.; Fontcuberta~i Morral,~A.; Stutzmann,~M. Surface passivation and
  self-regulated shell growth in selective area-grown GaN–(Al{,}Ga)N
  core–shell nanowires. \emph{Nanoscale} \textbf{2017}, \emph{9},
  7179--7188\relax
\mciteBstWouldAddEndPuncttrue
\mciteSetBstMidEndSepPunct{\mcitedefaultmidpunct}
{\mcitedefaultendpunct}{\mcitedefaultseppunct}\relax
\EndOfBibitem
\bibitem[Lew Yan~Voon and Willatzen(2011)Lew Yan~Voon, and
  Willatzen]{lewyanvoon_2011}
Lew Yan~Voon,~L.~C.; Willatzen,~M. Electromechanical phenomena in semiconductor
  nanostructures. \emph{J. Appl. Phys.} \textbf{2011}, \emph{109}, 031101\relax
\mciteBstWouldAddEndPuncttrue
\mciteSetBstMidEndSepPunct{\mcitedefaultmidpunct}
{\mcitedefaultendpunct}{\mcitedefaultseppunct}\relax
\EndOfBibitem
\bibitem[Geijselaers \latin{et~al.}(2018)Geijselaers, Lehmann, Dick, and
  Pistol]{Geijselaers2018}
Geijselaers,~I.; Lehmann,~S.; Dick,~K.~A.; Pistol,~M.-E. {Radial band bending
  at wurtzite{\textendash}zinc-blende{\textendash}{GaAs} interfaces}.
  \emph{Nano Futures} \textbf{2018}, \emph{2}, 035002\relax
\mciteBstWouldAddEndPuncttrue
\mciteSetBstMidEndSepPunct{\mcitedefaultmidpunct}
{\mcitedefaultendpunct}{\mcitedefaultseppunct}\relax
\EndOfBibitem
\bibitem[Logothetidis \latin{et~al.}(1996)Logothetidis, Alexandrou, and
  Kokkou]{Logothetidis1996}
Logothetidis,~S.; Alexandrou,~I.; Kokkou,~S. {Optimization of TiN thin film
  growth with in situ monitoring: the effect of bias voltage and nitrogen flow
  rate}. \emph{Surf. Coat. Technol.} \textbf{1996}, \emph{80}, 66\relax
\mciteBstWouldAddEndPuncttrue
\mciteSetBstMidEndSepPunct{\mcitedefaultmidpunct}
{\mcitedefaultendpunct}{\mcitedefaultseppunct}\relax
\EndOfBibitem
\bibitem[Zgrabik and Hu(2015)Zgrabik, and Hu]{Zgrabik2015}
Zgrabik,~C.~M.; Hu,~E.~L. {Optimization of sputtered titanium nitride as a
  tunable metal for plasmonic applications}. \emph{Opt. Mater. Express}
  \textbf{2015}, \emph{5}, 2786--2797\relax
\mciteBstWouldAddEndPuncttrue
\mciteSetBstMidEndSepPunct{\mcitedefaultmidpunct}
{\mcitedefaultendpunct}{\mcitedefaultseppunct}\relax
\EndOfBibitem
\bibitem[Suzuki and Hirabayashi(1993)Suzuki, and Hirabayashi]{Suzuki1993}
Suzuki,~T.; Hirabayashi,~Y. {First Observation of the
  Si(111)-$7\!\times\!7\!\leftrightarrow\!1\!\times\!1$ Phase Transition by the
  Optical Second Harmonic Generation}. \emph{Jpn. J. Appl. Phys.}
  \textbf{1993}, \emph{32}, L610--L613\relax
\mciteBstWouldAddEndPuncttrue
\mciteSetBstMidEndSepPunct{\mcitedefaultmidpunct}
{\mcitedefaultendpunct}{\mcitedefaultseppunct}\relax
\EndOfBibitem
\bibitem[Briggs \latin{et~al.}(2017)Briggs, Naik, Zhao, Petach, Sahasrabuddhe,
  Goldhaber-Gordon, Melosh, and Dionne]{Briggs2017}
Briggs,~J.~A.; Naik,~G.~V.; Zhao,~Y.; Petach,~T.~A.; Sahasrabuddhe,~K.;
  Goldhaber-Gordon,~D.; Melosh,~N.~A.; Dionne,~J.~A. {Temperature-dependent
  optical properties of titanium nitride}. \emph{Appl. Phys. Lett.}
  \textbf{2017}, \emph{110}, 101901\relax
\mciteBstWouldAddEndPuncttrue
\mciteSetBstMidEndSepPunct{\mcitedefaultmidpunct}
{\mcitedefaultendpunct}{\mcitedefaultseppunct}\relax
\EndOfBibitem
\end{mcitethebibliography}


\providecommand{\latin}[1]{#1}
\makeatletter
\providecommand{\doi}
  {\begingroup\let\do\@makeother\dospecials
  \catcode`\{=1 \catcode`\}=2 \doi@aux}
\providecommand{\doi@aux}[1]{\endgroup\texttt{#1}}
\makeatother
\providecommand*\mcitethebibliography{\thebibliography}
\csname @ifundefined\endcsname{endmcitethebibliography}
  {\let\endmcitethebibliography\endthebibliography}{}
\begin{mcitethebibliography}{13}
\providecommand*\natexlab[1]{#1}
\providecommand*\mciteSetBstSublistMode[1]{}
\providecommand*\mciteSetBstMaxWidthForm[2]{}
\providecommand*\mciteBstWouldAddEndPuncttrue
  {\def\EndOfBibitem{\unskip.}}
\providecommand*\mciteBstWouldAddEndPunctfalse
  {\let\EndOfBibitem\relax}
\providecommand*\mciteSetBstMidEndSepPunct[3]{}
\providecommand*\mciteSetBstSublistLabelBeginEnd[3]{}
\providecommand*\EndOfBibitem{}
\mciteSetBstSublistMode{f}
\mciteSetBstMaxWidthForm{subitem}{(\alph{mcitesubitemcount})}
\mciteSetBstSublistLabelBeginEnd
  {\mcitemaxwidthsubitemform\space}
  {\relax}
  {\relax}

\bibitem[Corfdir \latin{et~al.}(2014)Corfdir, Zettler, Hauswald,
  Fern\'andez-Garrido, Brandt, and Lefebvre]{Corfdir2014a}
Corfdir,~P.; Zettler,~J.~K.; Hauswald,~C.; Fern\'andez-Garrido,~S.; Brandt,~O.;
  Lefebvre,~P. {Sub-meV linewidth in {GaN} nanowire ensembles: Absence of
  surface excitons due to the field ionization of donors}. \emph{Phys. Rev. B}
  \textbf{2014}, \emph{90}, 205301\relax
\mciteBstWouldAddEndPuncttrue
\mciteSetBstMidEndSepPunct{\mcitedefaultmidpunct}
{\mcitedefaultendpunct}{\mcitedefaultseppunct}\relax
\EndOfBibitem
\bibitem[Calarco \latin{et~al.}(2005)Calarco, Marso, Richter, Aykanat, Meijers,
  {V D Hart}, Stoica, and L{\"{u}}th]{Calarco2005}
Calarco,~R.; Marso,~M.; Richter,~T.; Aykanat,~A.~I.; Meijers,~R.~J.; {V D
  Hart},~A.; Stoica,~T.; L{\"{u}}th,~H. {Size-dependent photoconductivity in
  {MBE}-grown {GaN}-nanowires.} \emph{Nano Lett.} \textbf{2005}, \emph{5},
  981--984\relax
\mciteBstWouldAddEndPuncttrue
\mciteSetBstMidEndSepPunct{\mcitedefaultmidpunct}
{\mcitedefaultendpunct}{\mcitedefaultseppunct}\relax
\EndOfBibitem
\bibitem[Yamabe \latin{et~al.}(1977)Yamabe, Tachibana, and
  Silverstone]{Yamabe1977}
Yamabe,~T.; Tachibana,~A.; Silverstone,~H.~J. Theory of the ionization of the
  hydrogen atom by an external electrostatic field. \emph{Phys. Rev. A}
  \textbf{1977}, \emph{16}, 877--890\relax
\mciteBstWouldAddEndPuncttrue
\mciteSetBstMidEndSepPunct{\mcitedefaultmidpunct}
{\mcitedefaultendpunct}{\mcitedefaultseppunct}\relax
\EndOfBibitem
\bibitem[Banavar \latin{et~al.}(1979)Banavar, Coon, and Derkits]{Banavar1979}
Banavar,~J.~R.; Coon,~D.~D.; Derkits,~G.~E. Low‐temperature field ionization
  of localized impurity levels in semiconductors. \emph{Appl. Phys. Lett.}
  \textbf{1979}, \emph{34}, 94--96\relax
\mciteBstWouldAddEndPuncttrue
\mciteSetBstMidEndSepPunct{\mcitedefaultmidpunct}
{\mcitedefaultendpunct}{\mcitedefaultseppunct}\relax
\EndOfBibitem
\bibitem[Kaganer \latin{et~al.}(2018)Kaganer, Sabelfeld, and
  Brandt]{Kaganer2018}
Kaganer,~V.~M.; Sabelfeld,~K.~K.; Brandt,~O. {Piezoelectric field, exciton
  lifetime, and cathodoluminescence intensity at threading dislocations in
  {GaN}\{0001\}}. \emph{Appl. Phys. Lett.} \textbf{2018}, \emph{112},
  122101\relax
\mciteBstWouldAddEndPuncttrue
\mciteSetBstMidEndSepPunct{\mcitedefaultmidpunct}
{\mcitedefaultendpunct}{\mcitedefaultseppunct}\relax
\EndOfBibitem
\bibitem[Adachi(1992)]{Adachi1992}
Adachi,~S. \emph{Physical properties of {III-V} Semiconductor compounds: {InP},
  InAs, {GaAs}, GaP, {InGaAs} and InGaAsP}; Willey Interscience, New York,
  1992\relax
\mciteBstWouldAddEndPuncttrue
\mciteSetBstMidEndSepPunct{\mcitedefaultmidpunct}
{\mcitedefaultendpunct}{\mcitedefaultseppunct}\relax
\EndOfBibitem
\bibitem[Vurgaftman \latin{et~al.}(2001)Vurgaftman, Meyer, and
  Ram-Mohan]{Vurgaftman2001}
Vurgaftman,~I.; Meyer,~J.~R.; Ram-Mohan,~L.~R. Band parameters for {III--V}
  compound semiconductors and their alloys. \emph{J. Appl. Phys.}
  \textbf{2001}, \emph{89}, 5815--5875\relax
\mciteBstWouldAddEndPuncttrue
\mciteSetBstMidEndSepPunct{\mcitedefaultmidpunct}
{\mcitedefaultendpunct}{\mcitedefaultseppunct}\relax
\EndOfBibitem
\bibitem[Meiners(1986)]{Meiners1986}
Meiners,~L.~G. Temperature dependence of the dielectric constant of {InP}.
  \emph{J. Appl. Phys.} \textbf{1986}, \emph{59}, 1611--1613\relax
\mciteBstWouldAddEndPuncttrue
\mciteSetBstMidEndSepPunct{\mcitedefaultmidpunct}
{\mcitedefaultendpunct}{\mcitedefaultseppunct}\relax
\EndOfBibitem
\bibitem[Winkelnkemper \latin{et~al.}(2006)Winkelnkemper, Schliwa, and
  Bimberg]{Winkelnkemper2006}
Winkelnkemper,~M.; Schliwa,~A.; Bimberg,~D. Interrelation of structural and
  electronic properties in {$In_{x}Ga_{1-x}N$}/{$GaN$} quantum dots using an
  eight-band $k \cdot p$ model. \emph{Phys. Rev. B} \textbf{2006}, \emph{74},
  155322\relax
\mciteBstWouldAddEndPuncttrue
\mciteSetBstMidEndSepPunct{\mcitedefaultmidpunct}
{\mcitedefaultendpunct}{\mcitedefaultseppunct}\relax
\EndOfBibitem
\bibitem[Syrbu \latin{et~al.}(2004)Syrbu, Tiginyanu, Zalamai, Ursaki, and
  Rusu]{Syrbu2004}
Syrbu,~N.; Tiginyanu,~I.; Zalamai,~V.; Ursaki,~V.; Rusu,~E. Exciton polariton
  spectra and carrier effective masses in {ZnO} single crystals. \emph{Physica
  B} \textbf{2004}, \emph{353}, 111--115\relax
\mciteBstWouldAddEndPuncttrue
\mciteSetBstMidEndSepPunct{\mcitedefaultmidpunct}
{\mcitedefaultendpunct}{\mcitedefaultseppunct}\relax
\EndOfBibitem
\bibitem[Oshikiri \latin{et~al.}(2001)Oshikiri, Imanaka, Aryasetiawan, and
  Kido]{Oshikiri2001}
Oshikiri,~M.; Imanaka,~Y.; Aryasetiawan,~F.; Kido,~G. Comparison of the
  electron effective mass of the n-type {ZnO} in the wurtzite structure
  measured by cyclotron resonance and calculated from first principle theory.
  \emph{Physica B} \textbf{2001}, \emph{298}, 472--476\relax
\mciteBstWouldAddEndPuncttrue
\mciteSetBstMidEndSepPunct{\mcitedefaultmidpunct}
{\mcitedefaultendpunct}{\mcitedefaultseppunct}\relax
\EndOfBibitem
\bibitem[Morhain \latin{et~al.}(2005)Morhain, Bretagnon, Lefebvre, Tang,
  Valvin, Guillet, Gil, Taliercio, Teisseire-Doninelli, Vinter, and
  Deparis]{Morhain2005}
Morhain,~C.; Bretagnon,~T.; Lefebvre,~P.; Tang,~X.; Valvin,~P.; Guillet,~T.;
  Gil,~B.; Taliercio,~T.; Teisseire-Doninelli,~M.; Vinter,~B.; Deparis,~C.
  Internal electric field in wurtzite {$ZnO$}/{$Zn_{0.78}Mg_{0.22}O$} quantum
  wells. \emph{Phys. Rev. B} \textbf{2005}, \emph{72}, 241305\relax
\mciteBstWouldAddEndPuncttrue
\mciteSetBstMidEndSepPunct{\mcitedefaultmidpunct}
{\mcitedefaultendpunct}{\mcitedefaultseppunct}\relax
\EndOfBibitem
\end{mcitethebibliography}

\end{document}


\section{Case of \emph{thin} nanowires}
Nanowires (NWs) are considered \emph{thin} if the diffusion time of the exciton along the NW radius is negligible compared to the nonradiative recombination time of the exciton at the NW surface. This condition is met for $d \ll L_\text{D}^*$, where $d$ denotes the NW diameter and $L_\text{D}^* = (D \tau_\text{i}^\text{surf})^{1/2}$ represents the diffusion length of the free exciton in the characteristic time $\tau_\text{i}^\text{surf}$ for exciton field ionization at the NW surface with $D$ denoting the free exciton diffusivity. For the NWs under scrutiny in the main text, the range of diameters satisfying the condition $d \ll L_\text{D}^*$ will be shown in section~\ref{sec:LDstar} to correspond to $d < 23$\,nm.

\subsection{Electric fields in NWs of arbitrary diameter}
Surface band bending occurs in a NW due to a charge transfer between surface and bulk states. The corresponding distribution of surface electric fields is calculated here for a NW of arbitrary diameter by solving the Poisson equation in cylindrical geometry and in the approximation of continuous and homogeneous charge distribution in the NW bulk.\footnote{Note that this approximation holds only above a certain NW diameter for any given doping density. Below this value, the electronic potential becomes nonparabolic and eventually entirely irregular due to the discrete and random position of donors (see Ref.~\citenum{Corfdir2014a} for details). However, the \emph{average} electric field in the NW still decreases with decreasing diameter, and our arguments thus remain valid even for NWs with nonparabolic potential.} The  NW is completely depleted once its diameter $d$ is below the critical value $d_\text{depl}$ amounting to:\cite{Calarco2005}
\begin{equation}
    d_\text{depl} = \sqrt{\frac{16 \varepsilon \varepsilon_\text{0} \Phi }{q N_\text{D}}}\,,
\end{equation}
with the elementary charge $q$, the donor concentration $N_\text{D}$, the permittivity $\varepsilon\varepsilon_\text{0}$ and the energy difference between the surface states and the conduction band minimum $\Phi$ [$\approx0.55$\,eV for oxidized GaN$(1\bar{1}00)$\cite{Calarco2005}].

\begin{figure*}
\includegraphics[width = .7\linewidth]{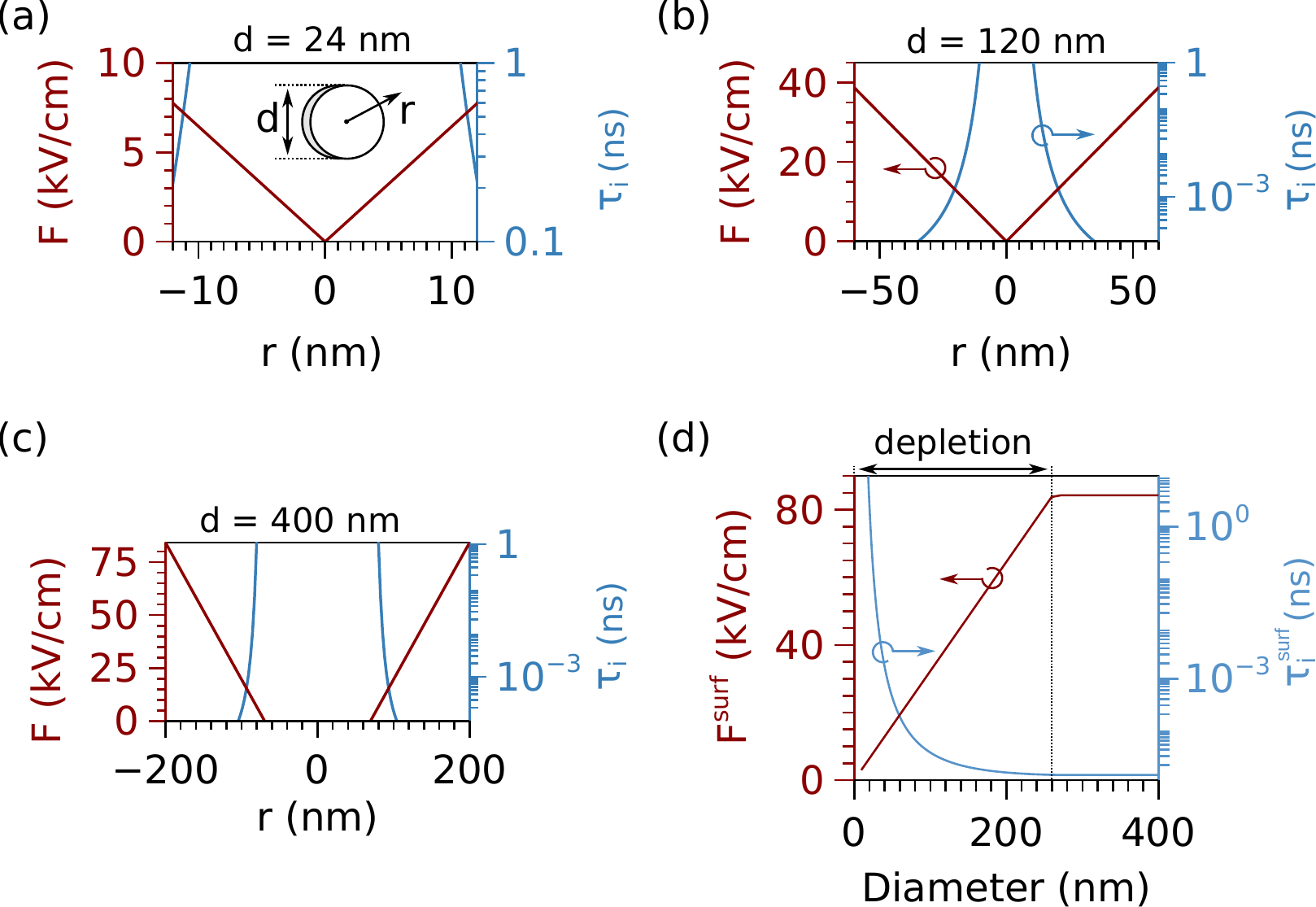}%
\caption{(a)-(c) Strength of the electric field $F$ and the related exciton ionization time $\tau_\text{i}$ plotted along the NW radius for three exemplary NW diameters. (d) Strength of the electric field $F^\text{surf}$ and the related exciton ionization time $\tau_\text{i}^\text{surf}$ at the NW surface given as function of the NW diameter. Electric fields are obtained from Eqs.\,\ref{eq:F} and \ref{eq:Fdepl} with $N_\text{D} = 7\times 10^{16}$\,cm$^{-3}$ and $\Phi = 0.55$\,eV resulting in $d_\text{depl} = 260$\,nm. The ionization time is obtained from Eq.\,\ref{eq:dissociation} using the material parameters of GaN given in Table\,\ref{tab:param}.
\label{fig:F}}
\end{figure*}

For $d > d_\text{depl}$, the strength of the surface band bending has no dependence on the NW diameter and the associated electric field $F(u)$, with $u = d/2 - r$, amounts to:\cite{Corfdir2014a}
\begin{align}
    &F(u = \frac{d}{2} -r) = \\
    &
 \begin{cases}
     (1-\dfrac{2u}{d_\text{depl}}) \dfrac{qN_\text{D} d_\text{depl}}{ 4\varepsilon \varepsilon_\text{0} } & \text{ for } u \in [0, \dfrac{d_\text{depl}}{2}] \\
     0 & \text{ for } u \in [\dfrac{d_\text{depl}}{2}, \dfrac{d}{2}] \,,\\
 \end{cases}
 \label{eq:F}
\end{align}
where $r$ is the radial position in cylindrical coordinates ($r=0$ in the NW center). The field strength reaches its maximum value at the surface:
\begin{equation}
    F^\text{surf} = \frac{qN_\text{D}d_\text{depl}}{4\varepsilon \varepsilon_\text{0}}\,.
\end{equation}
For $d < d_\text{depl}$, the strength of the surface band bending decreases quadratically with the NW diameter, and the associated electric field $F(r)$ amounts to:\cite{Corfdir2014a}
\begin{equation}
F(r) = \frac{q N_\text{D} r}{ 2\varepsilon \varepsilon_\text{0} }\,.
\label{eq:Fdepl}
\end{equation}
The field strength also reaches its maximum value at the surface:
\begin{equation}
    F^\text{surf}_\text{depl} = \frac{qN_\text{D}d}{4\varepsilon \varepsilon_\text{0}}\,.
\end{equation}
Using Eqs.\,\ref{eq:F} and \ref{eq:Fdepl} and taking a doping level $N_\text{D} = 7 \times 10^{16}$\,cm$^{-3}$ (for which $d_\text{depl} = 260$\,nm), the strength of the electric field calculated along the NW radius for three exemplary NW diameters ($d = 24$, $120$, and $400$\,nm) is shown in Figure~\ref{fig:F}(a)--\ref{fig:F}(c). In the two first cases, the NWs are entirely depleted, and the electric field vanishes only at the position $r = 0$. In the last case, depletion occurs only in the surface vicinity, and the NW core is free of electric fields. Figure~\ref{fig:F}(d) gives the strength of the electric field at the NW surface as a function of the NW diameter. For fully depleted NWs ($d < d_\text{depl}$), the surface electric field increases linearly with diameter. Once $d > d_\text{depl}$, the surface electric field saturates. 

\subsection{Field ionization of excitons}\label{sec:ionization}
The ionization rate $1/\tau_\text{i}$ for a free exciton subjected to an electric field $F$ is derived from the ionization probability (per unit time) of  the hydrogen atom\cite{Yamabe1977,Banavar1979} following Ref.\,\citenum{Kaganer2018}:
\begin{equation}\label{eq:dissociation}
    \frac{1}{\tau_\text{i}} = \omega \frac{4F_\text{0}}{F} \text{exp } \left(-\frac{2F_\text{0}}{3F}\right)\,,
\end{equation}
where the frequency $\omega$ and the electric field $F_\text{0}$ are given in terms of the exciton binding energy $E_\text{b}$ as $\omega = E_\text{b}/\hbar$ and $F_\text{0} = 2E_\text{b}/ q a_\text{b}$. Here, $\hbar$ denotes the reduced Planck constant and $a_\text{b}$ the Bohr radius. Values of $E_\text{b}$ and $a_\text{b}$ for several semiconductors commonly synthesized in the shape of NWs and relevant for optoelectronic applications are given in Table\,\ref{tab:param}. Using Eq.\,\ref{eq:dissociation}, we define the critical field $F_\text{crit}$ as the electric field required for ionization of the free exciton in about $1$\,ps. The calculated values reported in Table\,\ref{tab:param} evidence large variations between the different materials. The respective increase in the exciton binding energy for GaAs, InP, GaN, and ZnO significantly increases $F_\text{crit}$. This principally makes GaN more robust than GaAs and InP against nonradiative recombination mediated by electric fields. Yet, GaN is far surpassed by ZnO in this respect. 

\begin{table*}
   \caption{Critical field ($F_\text{crit}$), exciton Bohr radius ($a_\text{b} = 4 \pi \varepsilon \varepsilon_\text{0} \hbar^2 / \mu q^2$) and binding energy\\($E_\text{b}=\hbar^2 / 2 \mu a_\text{b}^2$) for several zincblende (ZB) and wurtzite (WZ) semiconductors with the permittivity $\varepsilon \varepsilon_0$, the elementary charge $q$, and the reduced mass $\mu$ of the exciton [$\mu = m_\text{e}m_\text{h}/(m_\text{e} + m_\text{h})$ with $m_\text{e}$ and $m_\text{h}$ denoting the masses of the electron and hole, respectively].  
    \label{tab:param}}
    \begin{tabular}{ccccccc}\hline\hline
        &$F_\text{crit}$ (kV/cm) &   $a_\text{b}$ (nm)                     &   $E_\text{b}$ (meV)   & $\varepsilon$        & $m_\text{e}$ & $m_\text{h}$ \\\hline
               GaAs (ZB)     &$1.0$&   $12$     &   $4.4$   & $13.18$ \cite{Adachi1992}      & $0.067$  \cite{Vurgaftman2001}  & $0.35$  \cite{Vurgaftman2001}   \\            
               InP (ZB)       &$1.7$&   $9.1$    &   $6.7$   & $11.76$ \cite{Meiners1986}     &$0.079$  \cite{Vurgaftman2001}  &  $0.53$  \cite{Vurgaftman2001} \\
               GaN  (WZ)     &$14$&   $3.1$    &   $24$    & $9.8$ \cite{Winkelnkemper2006} & $0.2$  \cite{Vurgaftman2001}    & $1$  \cite{Vurgaftman2001}\\
               ZnO (WZ)      &$52$&   $1.8$    &   $61$    & $6.4$ \cite{Syrbu2004}        & $0.24$  \cite{Oshikiri2001}     & $0.78$  \cite{Morhain2005}    \\     
    \end{tabular}
    
\end{table*}

\begin{figure}
\includegraphics[width = \linewidth]{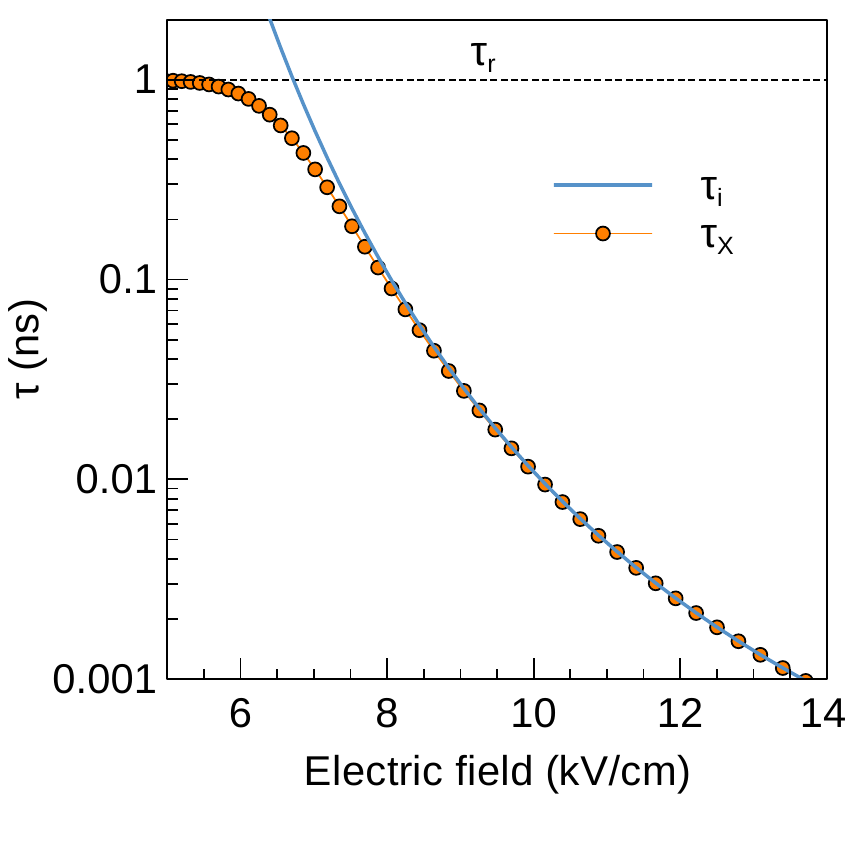}%
\caption{Field ionization time  $\tau_\text{i}$ of the exciton and its effective lifetime $\tau_\text{X}$ in GaN as function of the electric field strength assuming a radiative decay $\tau_\text{r} = 1$\,ns and no other competing decay channels.
\label{fig:tauE}}
\end{figure}

In Figure~\ref{fig:tauE}, the field ionization time $\tau_\text{i}$ of the free exciton in GaN and its effective lifetime $\tau_\text{X}$ ($\tau_\text{X}^{-1} = \tau_\text{i}^{-1} + \tau_\text{r}^{-1}$) are plotted as a function of the electric field strength. $\tau_\text{i}$ exhibits an abrupt dependence on the electric field, reaching $1$\,ns at $7$\,kV/cm and already $1$\,ps at $14$\,kV/cm. Hence, below $6$\,kV/cm, electric fields have a negligible impact on the exciton recombination rate. 

The characteristic time for field ionization of the exciton following from Eq.\,\ref{eq:dissociation} is shown in Figures~\ref{fig:F}(a)--\ref{fig:F}(d). In the NW core, the dissociation rate is negligible compared to the radiative recombination ($\tau_\text{i} \gg \tau_\text{r} \approx 1$\,ns), whereas it becomes significant when getting closer to the surface, especially for thick NWs. 

\subsection{Exciton decay}
As sketched in panel (ii) of Figure~3(b) in the main text, the decay time $\tau_< $ of the $(D^0,X_\text{A})$ exciton in \emph{thin} NWs encompasses the radiative decay in a characteristic time $\tau_\text{r} = 1$\,ns and the  nonradiative decay via field ionization of the free exciton in the surface electric field. The shortest ionization time that the free exciton can experience in the NW is considered as the rate-limiting step for the nonradiative decay, since the small diameter of the NW ensures a negligible diffusion time of the free exciton (in agreement with $d \ll L_\text{D}^*$). The nonradiative decay time thus equals the field ionization time at the NW surface $\tau_{i}^\text{surf}$, and the decay of the $(D^0,X_\text{A})$ exciton eventually amounts to:
\begin{equation}
    \tau_< = \left(1/ \tau_\text{r} + 1/ \tau_\text{i}^\text{surf} \right)^{-1}\,.
\end{equation}

\section{Case of \emph{thick} NWs}
\emph{Thick} NWs are characterized by $d >d_\text{DL}$, where $d_\text{DL}$ is the diameter above which an exciton dead layer develops below the surface. Here, for the NWs under scrutiny, $d_\text{DL}$ will be shown to amount to $40$\,nm. 

\subsection{Exciton dead layer}
\begin{figure*}
\includegraphics[width = .7\linewidth]{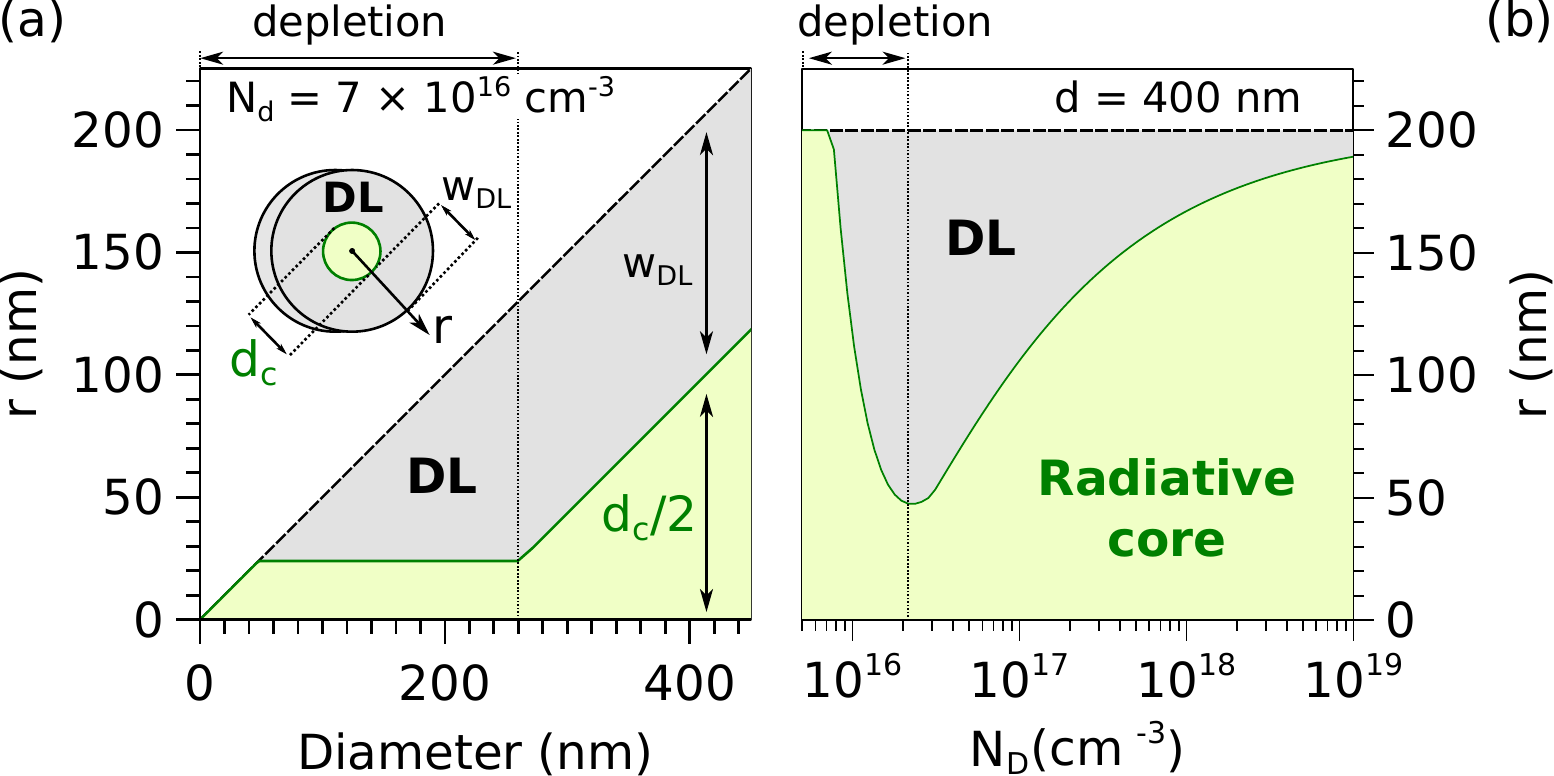}%
\caption{Radial width of the NW radiative core $d_\text{c}$ and of the dead layer $w_\text{DL}$ as a function (a) of the NW diameter and (b) of the doping level.
\label{fig:dead_layer}}
\end{figure*}

A close examination of Figure~\ref{fig:F} shows that above a certain diameter a sub-surface region with electric fields exceeding $F\geq F_\text{crit} = 14$\,kV/cm develops. This region is described as a dead layer since the nonradiative decay by field ionization is much faster than the radiative one ($\tau_\text{i} \ll \tau_\text{r} = 1$\,ns). The radial thickness $w_\text{DL}$ of this layer satisfies the equation:
\begin{equation}\label{eq:wDL}
    F(w_\text{DL}) = (1 - \frac{2 w_\text{DL}}{d_\text{depl}}) \frac{qN_\text{D}d_\text{depl}}{4 \varepsilon \varepsilon_\text{0}} = 14 \text{ kV/cm}\,,
\end{equation}
and the diameter of the radiative core of the NW $d_\text{c}$ can be determined correspondingly as:
\begin{equation}\label{eq:dcore}
    d_\text{c} = d - 2w_\text{DL}\,.
\end{equation}
Figures~\ref{fig:dead_layer}(a) and \ref{fig:dead_layer}(b) indicate how the NW volume splits off between the dead layer and the radiative core for NWs with different diameters and doping level, respectively. A reduction of the dead layer volume relative to the NW volume is obtained only by increasing the NW diameter or by increasing (decreasing) the doping level for nondepleted (depleted) NWs. Yet, the dead layer completely disappears from the NW only in the case of thin NWs and in the absence of doping. Figure~\ref{fig:dead_layer}(a) further illustrates a peculiarity of depleted NWs, namely, the constant diameter of the radiative core regardless of the actual NW diameter. 

\subsection{Exciton decay}
\begin{figure}
\includegraphics[width = \linewidth]{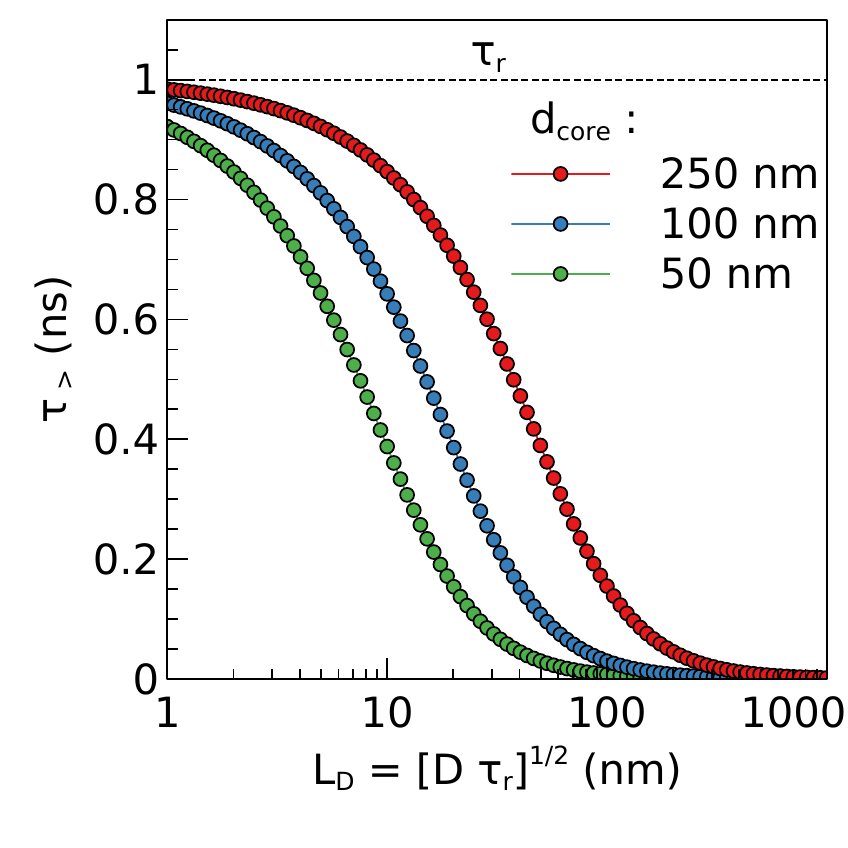}%
\caption{Decay time $\tau_>$ of the $(D^0,X_\text{A})$ exciton in \emph{thick} NWs as a function of the diffusion length $L_\text{D}$ of the free exciton for three exemplary NW core diameter. $\tau_{d > \tau_\text{DL}}$ is given by Eq.\,\ref{eq:tau_diff}.
\label{fig:tau_diff}}
\end{figure}

As sketched in panel (ii) of Figure~3(c) in the main text, the decay time $\tau_>$ of the $(D^0,X_\text{A})$ exciton in \emph{thick} NWs encompasses the radiative decay in a characteristic time $\tau_\text{r} = 1$\,ns and the  nonradiative decay via field ionization of the free exciton in the dead layer. Since the recombination velocity in the dead layer is infinite, the diffusion time of the exciton from the NW core to the edge of the dead layer is now the rate-limiting step for the nonradiative decay, in agreement with $d > d_\text{DL} \Leftrightarrow d  \gg L_\text{D}^*$. In the approximation of a uniform creation of excitons within the NW core, the density of free excitons $n_\text{X}$ thus satisfies the equation:
\begin{equation}
    \begin{aligned}
     D \vec \nabla^2 n_\text{X} + G - \frac{n_\text{X}}{\tau_\text{r}} &= 0\,,\\
    \end{aligned}
 \label{eq:diff}
\end{equation}
where $D$ is the exciton diffusivity, $G$ the generation rate and $\tau_\text{r}$ the radiative lifetime. In cylindrical coordinates ($r = 0$ in the NW center), the boundary conditions are $|\vec \nabla n_\text{X}| = 0$ at $r = 0$ and $n_\text{X} = 0$ at $r = d_\text{c}/2$. The first results from the cylindrical symmetry and the second accounts for the infinite recombination velocity at the edge of the dead layer. The integrated photoluminescence signal amounts to:
\begin{equation}
    I = \int_{V_\text{c}} \frac{n_\text{X}}{\tau_\text{r}}\,,
\end{equation}
where $V_\text{c}$ is the volume of the NW core. The radiative efficiency of the NW core is defined as
\begin{equation}
    \eta = \frac{I}{V_\text{c}G} = \frac{\tau_\text{nr}}{\tau_\text{r} + \tau_\text{nr}}\,,
\end{equation}
where $\tau_\text{nr}$ is the nonradiative decay time. Solving the equations gives the effective decay time of the exciton in the NW core:  
\begin{equation}
    \begin{aligned}\label{eq:tau_diff}
        \tau_> & = \left (\tau_\text{r}^{-1} + \tau_\text{nr}^{-1} \right )^{-1}\\
                                & = \tau_\text{r} \eta\\
                                & = \tau_\text{r} \frac{ I_\text{2}(d_\text{c}/2 L_\text{D} ) }{I_\text{0}(d_\text{c}/2 L_\text{D} )}\,,                             
    \end{aligned}
\end{equation}
where $I_\text{n}$ is the modified Bessel function of the first order and $L_\text{D} = (D\tau_\text{r})^{1/2}$ the free exciton diffusion length with $D$ the free exciton diffusivity. Figure~\ref{fig:tau_diff} gives the decay time $\tau_>$ of the $(D^0,X_\text{A})$ exciton in \emph{thick} NWs as function of the diffusion length $L_\text{D}$. An increase of the diffusion length is seen to reduce the recombination time of the $(D^0,X_\text{A})$ exciton, which leads to the counterintuitive fact that an enhanced material purity may actually result in an acceleration of the nonradiative decay in these NWs. 

\section{Delimiting \emph{thin} and \emph{thick} NWs}
\label{sec:LDstar}
\begin{figure}
\includegraphics[width = \linewidth]{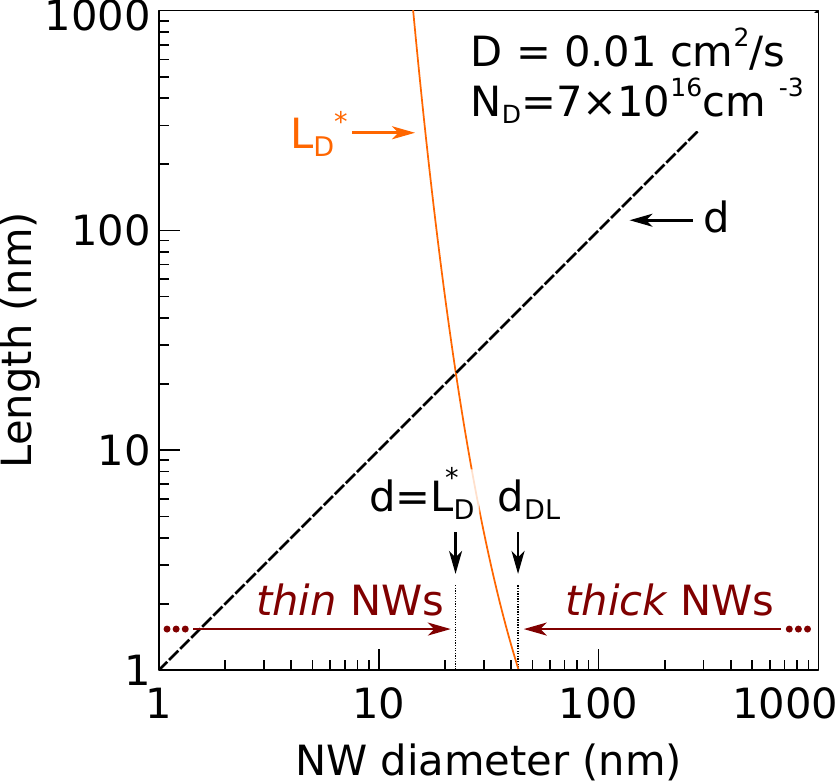}%
\caption{$L_\text{D}^* = \left( D\tau_\text{i}^\text{surf} \right) ^{1/2} $ calculated as function of the NW diameter for a doping concentration of $N_\text{d} = 7\times 10^{16}$\,cm$^{-3}$ and an exciton diffusivity $D = 0.01$\,cm$^2$/s. The critical diameters for which $d= L_\text{D}^*$ and $d=d_\text{DL}$ delimit the case of \emph{thin} and \emph{thick} NWs. 
\label{fig:LDstar}}
\end{figure}

\emph{Thin} and \emph{thick} NWs are respectively defined as $d \ll L_\text{D}^*$ and $d > d_\text{DL}$, where $L_\text{D}^* = (D \tau_\text{i}^\text{surf})^{1/2}$. As seen in the Figure 2(c) of the main text, a good match is obtained between the analytical expressions $\tau_<$ and $\tau_>$ and the exciton decay-time measured in as-grown and unintentionally doped GaN NWs when taking $N_\text{D} = 7 \times 10^{16}$\,cm$^{-3}$, $\Phi = 0.55$\,eV, $L_\text{D} =30$\,nm, $\tau_\text{r} = 1$\,ns, and $F_\text{crit} = 14$\,kV/cm. For these parameters, $d_\text{DL} = 40$\,nm and $d \ll L_\text{D}^* \Leftrightarrow d < 23$\,nm. Since $L_\text{D}^*$ depends on $d$, the latter equivalence is obtained graphically, as shown in Figure~\ref{fig:LDstar}. These boundaries would evidently differ for NWs with different surface states, doping level, or simply made of a different material. 

\bibliography{GaNNWs_Ti-TiN}